\documentclass[prd,showpacs,preprintnumbers,twocolumn,footinbib,amsmath,showkeys]{revtex4}
 \usepackage[dvips,final]{graphicx}
  \usepackage{amssymb,amsmath,epsfig,bm,pifont}
   \graphicspath{{./figs/}}
\usepackage{color}

\usepackage{relsize}
\newcommand{\babar}{{\mbox{\slshape B\kern-0.1em{\smaller A}\kern-0.1em
            B\kern-0.1em{\smaller A\kern-0.2em R}}}
           }
\usepackage[utf8]{inputenc}
\def\MSbar{\relax\ifmmode\overline
            {\rm MS}\else{$\overline{\rm MS}${ }}\fi}
\begin{document}
\thispagestyle{empty}
 \date{\today}
  \preprint{\hbox{RUB-TPII-01/2019}}

\title{Pion-photon transition form factor in QCD. \\
       Theoretical predictions and
       topology-based data analysis
       \\}
\author{N.~G.~Stefanis}
\email{stefanis@tp2.ruhr-uni-bochum.de}
\affiliation{Ruhr-Universit\"{a}t Bochum,
             Fakult\"{a}t f\"{u}r Physik and Astronomie,
             Institut f\"{u}r Theoretische Physik II,
             D-44780 Bochum, Germany\\}

\begin{abstract}
We discuss the evaluation of the transition form factor (TFF)
$F^{\gamma*\gamma\pi^0}(Q^2)$
by means of QCD theory and by state-space reconstruction from
topological data analysis.
We first calculate this quantity in terms of quark-gluon interactions
using light cone sum rules (LCSRs).
The spectral density includes radiative corrections in leading,
next-to-leading, and next-to-next-to-leading-order of perturbative QCD.
Besides, it takes into account the twist-four and twist-six terms.
The hard-scattering part in the LCSR is convoluted with various pion
distribution amplitudes with different morphologies in order to obtain
a wide range of predictions for the form factor, including two-loop
evolution which accounts for heavy-quark thresholds.
We then use nonlinear time series analysis to extract information on
the long-term $Q^2$ behavior of the measured scaled form factor in
terms of state-space attractors embedded in $\mathbb{R}^3$.
These are reconstructed by applying the Packard-Takens method of delays
to appropriate samplings of the data obtained in the CLEO,
\textit{BABAR}, and Belle single-tagged
$e^+e^- \rightarrow e^+e^-\pi^0$ experiments.
The corresponding lag plots show an aggregation of states around the
value
$Q^2F^{\gamma*\gamma\pi^0}(Q^2)\approx 0.165\pm 0.005$~GeV
pertaining to the momentum interval $Q^2\in [9-11]$~GeV$^2$.
We argue that this attractor portrait is a transient precursor of
a distribution of states peaking closer to the asymptotic limit
$Q^2F^{\gamma*\gamma\pi^0}(Q^2\to \infty)=\sqrt{2}f_\pi$~GeV.
More data with a regular increment of 1~GeV$^2$ in the range between 10
and 25~GeV$^2$ would be sufficient to faithfully determine the terminal
portrait of the attractor.
\\
\end{abstract}

\maketitle
\section{Introduction}
\label{sec:intro}
The Coulomb form factor of the neutral pion vanishes owing to
$C$-invariance.
But one can reveal the properties of the electromagnetic vertex of
$\pi^0$ in single-tagged $e^+e^-$ experiments by measuring the form
factor describing the process
$\pi^0\rightarrow\gamma^*\gamma$
in the spacelike region.
In this work, we consider this benchmark pion observable both
from the theoretical side by means of quantum chromodynamics (QCD)
and by nonlinear time series analysis of the data to reconstruct
the attractor of this observable in its state space.\footnote{An
attractor is operationally defined as the stable state (or set of
states) that a system (i.e., a point representing the system in its
state space) settles into, as it would be attracted towards that
state.}
The special importance of this transition form factor (TFF) derives
from the fact that in leading order of its microscopic description at
the level of quark and gluon degrees of freedom within QCD, it contains
only a single nonperturbative unknown, notably, the pion distribution
amplitude (DA).
Thus, it provides a handle to extract information about this
fundamental yet not directly measurable pion characteristic.
Besides, because of its simplicity, this observable offers the chance
to reveal the onset of QCD scaling related to asymptotic freedom from
the experimental data.

\begin{figure*}[ht]
\includegraphics[width=0.7\textwidth]{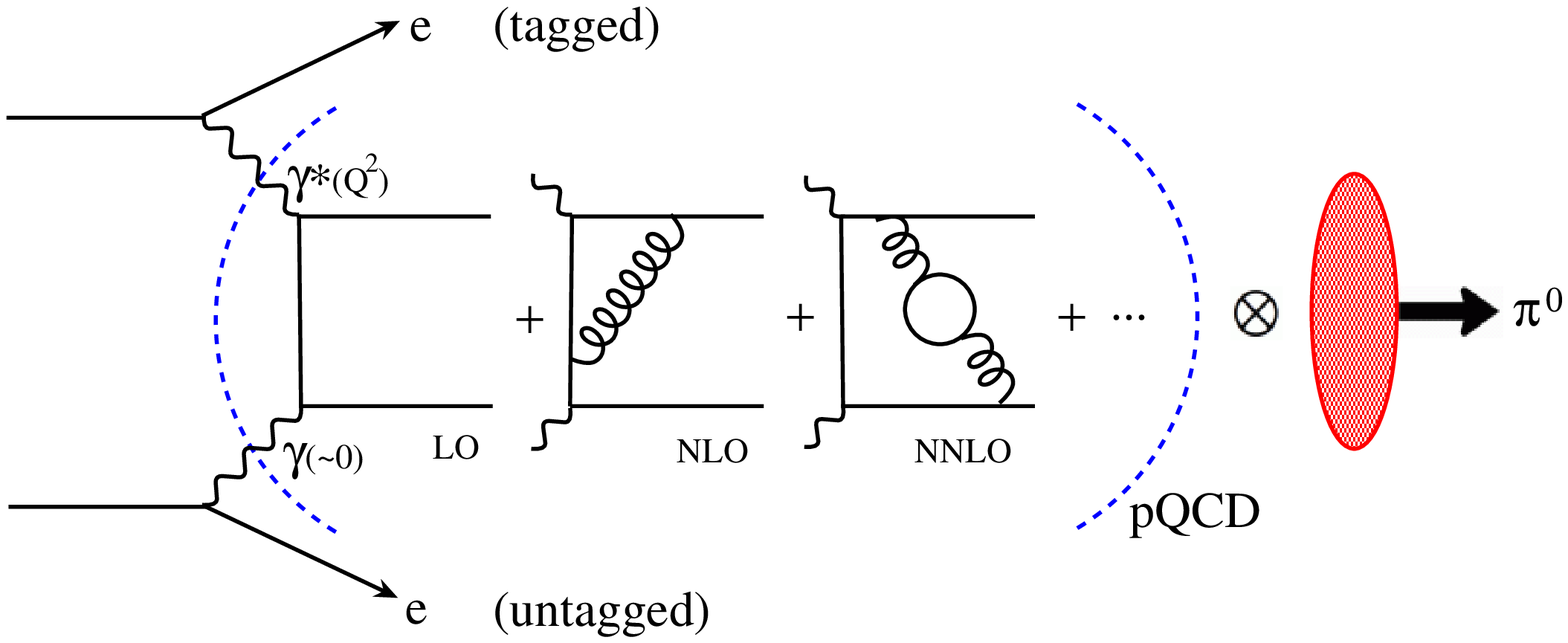}
\caption{Schematic illustration of the single-tag $\pi^0$ production
in a two-photon process with one photon having large virtuality $Q^2$,
while the other, emitted from the untagged electron (positron), is
almost real.
The transition form factor for the process
$\gamma^*\gamma\to q\bar{q}\to\pi^0$
is shown in factorized form in terms of a convolution
(denoted by the symbol $\bigotimes$)
which separates short- from long-distance dynamics.
The former contributes to the hard-scattering amplitude, which is
displayed in terms of some representative quark-gluon subprocesses
in leading (LO),
next-to-leading (NLO), and
next-to-next-to-leading order (NNLO)
of perturbative QCD (pQCD)
within the parenthesis denoted by a dashed blue line.
The solid (wavy) lines represent quarks (gluons).
The nonperturbative long-distance part is encoded in the universal
pion distribution amplitude denoted by the (red) shaded oval.
Higher-twist contributions and the large-distance structure of the
quasireal photon have been ignored here for simplicity; they are
discussed in the text.
\label{fig:2-photon-process}
}
\end{figure*}

The exclusive production of a pseudoscalar $\pi^0$ meson from
$e^{\pm}e^{\pm}$ scattering is described by the amplitude
\begin{equation}
 T_{\mu\nu}
=
  i \epsilon_{\mu\nu\alpha\beta}
  q_{1}^{\alpha} q_{2}^{\beta}
  F^{\gamma^{*}\gamma^{*}\pi^{0}}\left(q_{1}^{2}, q_{2}^{2}\right)
\label{eq:TFF-definition}
\end{equation}
in terms of the TFF
$F^{\gamma^{*}\gamma^{*}\pi^{0}}\left(q_{1}^{2}, q_{2}^{2}\right)$
and asymmetric kinematics $q_{1}^{2}\gg q_{2}^{2}$
with $q_{1}^{2}, q_{2}^{2} \gg m^2_\rho$ and $m_\rho=775$~MeV,
adjusted to a ``single-tagged'' experimental mode.
The highly off-the-mass-shell photon with virtuality
$Q^2\equiv-q_{1}^2=(p-p^\prime)^2$ is emitted from the tagged
electron (or positron), where $p$ and $p^\prime$ are the
four-momenta of the initial and final electrons (or positrons)
emerging at a finite relative angle.
The other photon has a very low virtuality $q^2\equiv -q_2^2\gtrsim 0$
because the momentum transfer to the untagged electron (or positron),
from which it is virtually emitted, is close to zero
(see Fig.\ \ref{fig:2-photon-process}).

Using a single-tagged experimental set-up, one measures the
differential cross section
$d\sigma(Q^2,q^2=0)/dQ^2$
for the above exclusive process by employing signal kinematics to
select events in which the $\pi^0$ and one final-state electron
(or positron) --- the ``tag'' --- are registered, while the other
lepton remains undetected.
Typically, one detects the decay products of the meson and either the
electron (or the positron) which emerges after the scattering at some
minimum angle relative to the $e^{+}e^{-}$ collision axis.
In this case, one virtual photon has a large spacelike momentum
(electron mass neglected)
\begin{equation}
  Q^2
\equiv
  -(p_\text{beam}-p_\text{tag})^2
=
  2E_\text{beam}E_\text{tag}(1-\cos\theta_\text{tag}) \, ,
\label{eq:large-mom}
\end{equation}
where $p_\text{beam}$ ($E_\text{beam}$) and
$p_\text{tag}$ ($E_\text{tag}$) are the four-momenta (energies)
of the incident beam-energy electron and the tag, respectively,
with $\theta_\text{tag}$ being the scattering angle,
while the other photon is almost real.
Then, the differential cross section
$d\sigma\left(e^+e^- \rightarrow e^+e^-R\right)/dQ^2$
can be linked to the TFF by
employing the Budnev, Ginzburg, Meledin, and Serbo (BGMS) formalism
\cite{Budnev:1974de}.
It states that the deviation of the production rate from the
expression describing point-like pions as $Q^2$ grows, amounts to a
measurement of the form factor.
These considerations apply also to other processes
$e^{+}e^{-}\to e^{+}e^{-}R$, where $R$ denotes one of the light
pseudoscalar mesons $\pi^0$, $\eta$, $\eta^\prime$.

For these mesons, there is only one scalar form factor,
which can be written in the form, see, e.g., \cite{Uehara:2012ag},
\begin{equation}
  |F^{\gamma^*\gamma R}(Q^2,q^2=0)|^2
=
  \frac{d\sigma(Q^2,q^2=0)/dQ^2}{2A(Q^2)} \, ,
\label{eq:TFF-cross-sect}
\end{equation}
where the quantity $A(Q^2)$ is calculable within quantum
electrodynamics (QED).
At zero momentum transfer, $Q^2=0$, one has by normalization
\begin{equation}
  |F^{\gamma^*\gamma R}(0,0)|^2
=
  \frac{1}{(4\pi\alpha)^2}
  \frac{64\pi \Gamma(R\rightarrow\gamma\gamma)}{M_{R}^3} \, ,
\label{eq:zero-mom}
\end{equation}
where $\alpha=1/137$ is the QED coupling constant,
$\Gamma(R\rightarrow\gamma\gamma)$
is the two-photon partial width of the meson $R$ and $M_{R}$ is its
mass.
These two quantities are known from other experiments.
The empirical form factor is extracted by comparing the measured
values of the cross section with those from a Monte Carlo (MC)
simulation,
\begin{equation}
  F^2(Q^2)
=
  \frac{(d\sigma/dQ^2)_\text{data}}{(d\sigma/dQ^2)_\text{MC}}
  F_\text{MC}^{2} \, ,
\label{eq_TFF-exp}
\end{equation}
where
$F_\text{MC}^{2}$ is a constant form factor.
Typically, the two-photon Monte Carlo program in single-tag experiments
is based on the BGMS formalism and describes the $Q^2$ (and $q^2$)
development of the TFF by the following approximate form based on the
factorization of the $Q^2$ and $q^2$ dependencies on account of vector
meson dominance \cite{Sakurai:1960ju}
(see, for instance, \cite{Gronberg:1997fj})
\begin{eqnarray}
  |F^{\gamma^*\gamma^* R}(Q^2,q^2)|^2
\! && \!\!\! =
  \frac{1}{(4\pi\alpha)^2}
  \frac{64\pi\Gamma(R\rightarrow \gamma\gamma)}{M_{R}^3}
\nonumber \\
&& \!\! \frac{1}{(1+Q^2/\Lambda_{R}^{2})^2}
   \frac{1}{(1+q^2/\Lambda_{R}^{2})^2} \, .
\label{eq:momentum-develop-TFF}
\end{eqnarray}
The pole-mass parameter $\Lambda_{R}\approx 770$~MeV is chosen to
reproduce the momentum-transfer dependence of the form factors.
Note that the calculated cross section $(d\sigma/dQ^2)_\text{MC}$
acquires a model-dependent uncertainty induced by the unknown
dependence on the momentum transfer to the untagged electron.

The behavior of the form factor is known theoretically in two limits.
At $Q^2\to 0$ and in the chiral limit of quark masses, one obtains
from the axial anomaly \cite{Adler:1969gk,Bell:1969ts}
\begin{equation}
  \lim_{Q^2\to 0}F^{\gamma^*\gamma\pi^0}(Q^2)
=
  \frac{1}{2\sqrt{2}\pi^2 f_{\pi}} \, ,
\label{eq:axial-anomaly}
\end{equation}
where $f_{\pi}=132$~MeV is the leptonic decay constant of the pion.
On the other hand, the asymptotic behavior of the form factor
at $Q^2\to\infty$ is determined from perturbative QCD to be
\cite{Lepage:1980fj,Brodsky:1981rp}
\begin{equation}
  \lim_{Q^2\to\infty} \mathcal{F}(Q^2)
  =
   \sqrt{2} f_\pi
  \approx 0.187~\mbox{GeV} \, ,
\label{eq:asy-TFF}
\end{equation}
where we used the convenient notation
\begin{equation}
  Q^2F^{\gamma^*\gamma\pi^0(Q^2)}\equiv\mathcal{F}(Q^2) \, .
\label{eq:scaled-TFF}
\end{equation}
The TFF in the $Q^2$ range between the aforementioned limits can be
phenomenologically described by the interpolation formula of Brodsky
and Lepage \cite{Brodsky:1981rp},
\begin{equation}
  F^{\gamma^*\gamma \pi}(Q^2)
=
  \frac{\sqrt{2}f_\pi}{4\pi^2f_\pi^2 + Q^2 } \ .
\label{eq:interpol}
\end{equation}
Analogous expressions hold for $\eta$ and $\eta^\prime$ with $f_\pi$
being replaced by the corresponding decay constants.

The asymptotic limit of the TFF at $Q^2\to \infty$ is the result of an
abstraction process that cannot be realized in real-world experiments.
To this end, one would need a \textit{long series} of high-precision
observations at higher and higher $Q^2$ values in order to minimize the
influence of statistical flukes.
Moreover, we don't know at which momentum transfer the TFF should
come close to the asymptotic limit.
We also have no clues whether this limit is approached uniformly from
below or from above; there is no sharp borderline between the
underlying dynamical regimes.
A selfconsistent calculation of the TFF within QCD encompasses various
regimes of dynamics from low $Q^2\lesssim1$~GeV$^2$, where perturbation
theory is unreliable and nonperturbative effects are eventually more
important but poorly known, up to high $Q^2$ values where one would
expect that the perturbative contributions, depicted in
Fig.\ \ref{fig:2-photon-process} in terms of quark-gluon diagrams,
prevail and provide an accurate dynamical picture within perturbative
QCD.
But again, there is no standard candle to fix a priori this crucial
crossing point between nonperturbative and perturbative physics
on route to collinear universality at $Q^2\to\infty$.
For a broad survey of strong-interaction dynamics at different momentum
scales, see \cite{Brambilla:2014jmp}.

In fact, there are mainly three different sources of nonperturbative
effects related to confinement that contribute to the TFF:
(i) mass generation due to Dynamical Chiral Symmetry Breaking (DCSB),
(ii) the bound-state dynamics of the pion,
and (iii) the hadronic content of the quasireal photon
that is emitted  from the untagged electron (or positron) at large
distances.
We do not address DCSB in this work, but we refer to other
approaches which account for this.
A reliable theoretical scheme able to include the other two
nonperturbative ingredients, together with higher-order perturbative
QCD contributions and nonperturbative higher-twist corrections, is
discussed in the next section.

Though such theoretical frameworks are extremely useful and embody
the high principles of QCD, one needs \emph{in practice} some
guide for organizing and systematizing the data directly at the
observational level, without appealing to underlying theoretical
explanations at the level of quarks and gluons.\footnote{Monte Carlo
simulations provide an effective tool to generate the physical
configurations of the system but cannot unveil the hidden dynamics.}
This is even more important in the case of contradictory experimental
data \cite{Aubert:2009mc,Uehara:2012ag} that are eventually indicating
discrepant observations applying to the same phenomenon; see
\cite{Bakulev:2012nh,Stefanis:2012yw} for a detailed comparison
of various theoretical approaches and a classification scheme of the
predictions.

To extract dynamical information from the existing data on the
experimental system, described by the quantity $\mathcal{F}(Q^2)$, we
employ for the first time in this field of physics a topology-based
data analysis in terms of nonlinear time-series
of measurements of this single scalar variable.
The aim is to reconstruct the dynamical evolution of
$\mathcal{F}(Q^2)$ in terms of delay-coordinate maps from these time
series that give rise to a trajectory representing
the entirety of the states of this quantity.
An introduction to the subject can be found in \cite{Williams:1997}.
To achieve this goal, we pursue the idea that the dynamical evolution
of the system in its state space asymptotically contracts onto an
unobservable attractor that can be reconstructed from scatter plots
of delayed time series extracted from the data
(measuring time in units of $Q^2$~[GeV$^2$]).
The key assumption is that some universal underlying determinism
exists (related to the natural flow of the dynamical system),
which may elude analysis using traditional methods in real space,
but shows up in the pattern of the data in the form of an attractor
in the lagged phase space of this
quantity.\footnote{The lagged phase space is a special case of a pseudo
phase space because in contrast to the standard phase space the axes
represent time-shifted values of the same variable.
The terms state space and pseudo (lagged) phase space are used here
interchangeably.}

To reconstruct the phase-space portrait of the attractor, the time
delay method pioneered by Packard et al. \cite{Packard:1980zz} and
independently by Takens \cite{Takens:1981} will be used.\footnote{The
first method employs for the attractor reconstruction time derivatives
formed from the data, while the second one involves time delays of the
measured quantity, providing the mathematical foundation for both
methods.}
The delay method is based on the existence theorem for the embedding of
manifolds in Euclidean spaces by Whitney \cite{10.2307/1968482} and
on Takens' embedding theorem for delay-coordinate maps that provides
a sufficient condition that the reconstructed attractor will have the
same dynamical properties as the unobservable attractor of the full
dynamical system of unknown dimension.
These works were followed by the publication of an algorithm
\cite{GRASSBERGER1983189} to compute the correlation dimension of the
reconstructed attractor, see \cite{Sauer:1991} for a mathematical
discussion of the embedding techniques and
\cite{Hegger:1998tis,Schreiber:1999sts,Pecora:2007,MARWAN2007237,Bradley:2015nts}
for technical reviews.
The concept of state-space reconstruction in terms of an attractor,
has been widely used across a range of disciplines dealing with data
that are generated sequentially in time, see \cite{Stam:2005}.

The objective in this work is to identify a data-driven long-term
pattern in the state-space of the physical TFF $\mathcal{F}(Q^2)$
and use regular samplings of the existing experimental data to
reconstruct an attractor in its state space.
If the phase-state portrait of the reconstructed attractor exhibits
a region of densely recurrent states in the vicinity of the asymptotic
TFF  value, given by Eq.\ (\ref{eq:asy-TFF}), then this state
aggregation quantified in terms of histograms,
can be interpreted as a clear indication for a saturating
behavior of the form factor in accordance with QCD.
Such a phase-space configuration of the hypothesized attractor cannot
be determined conclusively at present because, as we show in a later
chapter, the number of the elements of the lagged time series,
extracted from the available TFF data sets, is rather small at high
$Q^2$.
Nevertheless, replicating, grosso modo, the large-$Q^2$ QCD limit of
the TFF by an accurate phase portrait of the quasi-asymptotic attractor
directly from the data would not only provide a deeper theoretical
insight how the onset of QCD scaling is approached, it would
also deduce practical consequences for the data processing of future
experiments.
In fact, the main challenge in pursuing this methodology is the limited
data coverage of the $10-25$~GeV$^2$ region, requiring a more dense
data acquisition in steps of 1~GeV$^2$.
Here the planned Belle-II experiment could contribute significantly
in a momentum regime that offers a much better data-taking feasibility
than the measurements at much higher $Q^2$ values.
This option was barely explored in previous experiments.

The paper has two main parts.
In the first part (Sec.\ \ref{sec:analysis-theory}), we present
theoretical predictions obtained within QCD using the method of
light cone sum rules (LCSRs) \cite{Balitsky:1989ry,Khodjamirian:1997tk}
and fixed-order perturbation theory (FOPT).
The spectral density at the twist two level includes all presently
known radiative corrections up to the next-to-next-to-leading
order \cite{Mikhailov:2016klg}.
On the other hand, the twist-four term and the twist-six contribution
\cite{Agaev:2010aq} are also included.
Note that the radiative corrections to the TFF can be taken into
account in a resummed way by combining the LCSR method  with the
solution of the renormalization-group equation \cite{Ayala:2018ifo}.
Because in the present investigation we are mainly interested in the
behavior of $\mathcal{F}(Q^2)$ in the far-end $Q^2$ regime, where
these effects play a minor role, we use for simplicity FOPT.

As nonperturbative input, we employ two types of endpoint-suppressed
twist-two pion DAs: a family of bimodal DAs \cite{Bakulev:2001pa} and
a platykurtic DA determined more recently in
\cite{Stefanis:2014nla} and further discussed in
\cite{Stefanis:2014yha,Stefanis:2015qha}.
Both types of DAs are obtained from QCD sum rules with nonlocal
condensates (NLC).
The latter have been introduced long ago in
\cite{Mikhailov:1986be,Mikhailov:1988nz,Mikhailiov:1989mk,Bakulev:1991ps,Mikhailov:1991pt}.
The mentioned DAs serve in the present analysis as a reference point to
compare with TFF predictions obtained with external pion DAs within the
same LCSR-based scheme.
More explicitly, we derive predictions by employing the pion DAs
determined with the help of Dyson-Schwinger equations (DSE)
\cite{Chang:2013pq,Raya:2015gva},
a light-front (LF) model \cite{Choi:2014ifm},
a nonlocal chiral quark model from the instanton vacuum
\cite{Nam:2006sx},
and from holographic AdS/QCD \cite{Brodsky:2011yv,Brodsky:2011xx}.
To obtain the TFF at the measured $Q^2$ values, we use
Efremov-Radyushkin-Brodsky-Lepage (ERBL)
\cite{Efremov:1978rn,Lepage:1980fj} evolution at the NLO, i.e.,
two-loop level, which includes heavy-quark thresholds
(``global QCD scheme'' \cite{Shirkov:1994td}).
All calculated predictions are compared with the available data from
the CELLO \cite{Behrend:1990sr},
CLEO \cite{Gronberg:1997fj},
\textit{BABAR} \cite{Aubert:2009mc},
and Belle \cite{Uehara:2012ag} experiments.
The preliminary data of the BESIII Collaboration
\cite{Redmer:2018uew} have also been included.

The second part of the paper is presented in
Sec.\ \ref{sec:data-analysis}
and is devoted to the analysis of the data from the last three
mentioned experiments in the context of the state-space attractor
reconstruction.
The presentation includes the mathematical basis of the method and the
techniques for its application in the present analysis.
The key advantages of the reconstructed attractor in extracting
information from the data on the dynamics of the $\pi-\gamma$
transition process are worked out.
More importantly, it is shown that the phase-space attractor can
provide a shortcut to uncover the asymptotic properties of the TFF
from experiment at much lower $Q^2$ values than isolated
measurements at much higher momenta.
In Sec.\ \ref{sec:pred-forecasts}, we examine and discuss more
closely and critically the results obtained in the previous two main
sections comparing them with estimates from lattice QCD.
We summarize and conclude our analysis in Sec.\ \ref{sec:concl}.
The employed NLO evolution scheme with threshold inclusion is
considered in Appendix \ref{sec:global-NLO-evolution}.
The experimental data are displayed in Appendix \ref{sec:data-theory}
together with the values of the TFF and its chief uncertainties for the
Bakulev-Mikhailov-Stefanis (BMS) set of DAs \cite{Bakulev:2001pa}.
The analogous results for the platykurtic pion DA
\cite{Stefanis:2014nla} are also included, employing in both cases
the NLO evolution scheme worked out in the previous Appendix.
These numerical values update all our previously published results.

\section{Theoretical analysis of $Q^2F^{\gamma^{*}\gamma\pi^{0}}(Q^2)$
using LCSRs in QCD}
\label{sec:analysis-theory}
In this section we consider the calculation of the pion-photon TFF
using LCSRs \cite{Balitsky:1989ry,Khodjamirian:1997tk} in QCD
in conjunction with various pion DAs of twist two.
We first expose the applied formalism
\cite{Bakulev:2002hk,Mikhailov:2009kf,Mikhailov:2016klg}
and then continue with the presentation and discussion of the obtained
TFF predictions.

\subsection{LCSR approach to the pion-photon transition form factor}
\label{subsec:LCSR}
In QCD, the amplitude $T_{\mu\nu}$ describing the process
$\gamma^{*}(q_{1})\gamma^{*}(q_{2})\rightarrow \pi^{0}(P)$
(cf.\ (\ref{eq:TFF-definition}))
is defined by the correlation function
\begin{eqnarray}
&& \int\! d^{4}z\,e^{-iq_{1}\cdot z}
  \langle
         \pi^0 (P)| T\{j_\mu(z) j_\nu(0)\}| 0
  \rangle
=
  i\epsilon_{\mu\nu\alpha\beta}
  q_{1}^{\alpha} q_{2}^{\beta}
\nonumber \\
&&  ~~~~~~~~~~~~~~ \times ~
  F^{\gamma^{*}\gamma^{*}\pi^0}(Q^2,q^2)\ ,
\label{eq:matrix-element}
\end{eqnarray}
where
$j_\mu=\frac{2}{3}\bar{u}\gamma_\mu u - \frac{1}{3}\bar{d}\gamma_\mu d$
is the quark electromagnetic current.
Expanding the T-product of the composite (local) current operators in
terms of $Q^2$ and $q^2$
(assuming that they are both sufficiently large, i.e.,
$Q^2,~ q^2\gg \Lambda_{R}^{2}$, cf.\ (\ref{eq:momentum-develop-TFF})),
one gets by virtue of the factorization theorem, the LO term
\begin{equation}
  F^{\gamma^{*}\gamma^{*}\pi}(Q^{2},q^{2})
=
  N_\text{T} 
  \int_{0}^{1} dx
  \frac{1}{Q^{2}\bar{x} + q^{2}x}~\varphi_{\pi}^{(2)}(x)
\label{eq:leading-FF-term}
\end{equation}
with $N_\text{T}=\sqrt{2}f_{\pi}/3$ and $\varphi_{\pi}^{(2)}$ denoting
the pion DA of twist two.
For vanishing $q^2$ it reduces to the expression
\begin{eqnarray}
  \frac{3}{\sqrt{2}f_{\pi}}Q^2F_{\gamma^*\gamma\pi^0}^{(\rm LO)}(Q^2)
& = &
  \int_{0}^{1} \varphi_{\pi}^{(2)}(x)/x
=
  \langle 1/x\rangle_{\pi}
\nonumber \\
& = &
  3(1+a_{2}+a_{4}+a_{6}+ \ldots) \, ,
\nonumber \\
\label{eq:inv-mom}
\end{eqnarray}
where we recast the inverse moment $\langle 1/x\rangle_{\pi}$ in terms
of the projection coefficients $a_n$ on the set $\{\psi_n\}$
of the eigenfunctions of the one-loop ERBL evolution equation,
\begin{equation}
  \varphi_{\pi}^{(2)}(x,\mu^2)
=
  \psi_{0}(x) + \sum_{n=2,4,\ldots}^{\infty} a_{n}(\mu^2) \psi_{n}(x) \, .
\label{eq:DA-conf-exp}
\end{equation}
Here
$\psi_{0}(x)=6x(1-x)\equiv 6x\bar{x}$
is the asymptotic pion DA $\varphi_{\pi}^\text{asy}$ and the higher
eigenfunctions are given in terms of the Gegenbauer polynomials
$\psi_n(x)=6x\bar{x}C_{n}^{(3/2)}(x-\bar{x})$.

The pion DA parameterizes the matrix element
\begin{eqnarray}
  \langle 0| \bar{d}(z) \gamma_\mu\gamma_5 [z,0] u(0)
           | \pi(P)
  \rangle|_{z^{2}=0}
&& \!\!\!\!\! =
  if_\pi P_\mu \int_{0}^{1} dx e^{i x (z\cdot P)}
\nonumber \\
&& \times
  \varphi_{\pi}^{(2)} \left(x,\mu^2\right) \, ,
\label{eq:pion-DA}
\end{eqnarray}
where the path-ordered exponential (the lightlike gauge link)
$
 [z,0]
=
 \mathcal{P}\exp \left[
                       ig \int_{0}^{z} t_{a}A_{a}^{\mu}(y)dy_{\mu}
                 \right]
$
ensures gauge invariance.
It is set equal to unity by virtue of the light-cone gauge
$z\cdot A=0$.
Higher-twist DAs in the light cone operator product expansion of the
correlation function in (\ref{eq:matrix-element}) give contributions
to the TFF that are suppressed by inverse powers of $Q^2$.
Physically, $\varphi_{\pi}^{(2)}(x,Q^2)$ describes the partition of
the pion's longitudinal momentum between its two valence partons,
i.e., the quark and the antiquark,
with longitudinal-momentum fractions
$x_q=x=(k^0+k^3)/(P^0+P^3)=k^+/P^+$
and
$x_{\bar{q}}=1-x\equiv \bar{x}$, respectively.
It is normalized to unity,
$
 \int_{0}^{1}dx \varphi_{\pi}^{(2)}(x)=1
$, so that $a_0=1$.

The expansion coefficients $a_{n}(\mu^2)$ are hadronic parameters and
have to be determined nonperturbatively at the initial scale of
evolution $\mu^2$, but have a logarithmic $Q^2$ development via
$\alpha_s(Q^2)$ governed by the ERBL evolution equation,
see, for instance, \cite{Stefanis:1999wy} for a technical review.
The one-loop anomalous dimensions $\gamma_{n}^{(0)}$ are the
eigenvalues of $\psi_n(x)$ and are known in closed
form \cite{Lepage:1980fj}.
The ERBL evolution of the pion DA at the two-loop order is more
complicated because the matrix of the anomalous dimensions is
triangular in the $\{\psi_n(x)\}$ basis and contains off-diagonal
mixing coefficients
\cite{Dittes:1981aw,Sarmadi:1982yg,Mikhailov:1984ii,Mueller:1993hg,%
Mueller:1994cn,Bakulev:2002uc,Bakulev:2005vw,Agaev:2010aq}.
To obtain the TFF predictions in the present work, we employ a
two-loop evolution scheme, which updates the procedure given in
Appendix D of \cite{Bakulev:2002uc} by including the effects of
crossing heavy-quark thresholds
in the NLO anomalous dimensions $\gamma_{n}^{(1)}$ and also
in the evolution of the strong coupling,
see, e.g., \cite{Shirkov:1994td,Bakulev:2012sm,Ayala:2014pha}
and App.\ \ref{sec:global-NLO-evolution}.

Elevating the convolution form (\ref{eq:leading-FF-term}) to higher
orders, one has at the leading twist-two level
\cite{Efremov:1978rn,Lepage:1980fj}
\begin{eqnarray}
  F_\text{QCD}^{\gamma^{*}\gamma^{*}\pi^0}\!\!\left(Q^2,q^2,\mu_{\rm F}^2\right)
= && \!\!\!\!\!
  N_\text{T} 
  \int_{0}^{1} dx \,
  T\left(Q^2,q^2;\mu_{\rm F}^2;x\right)
\nonumber \\
\!\!\! && \!\!\! \times
  \varphi_{\pi}^{(2)}\left(x,\mu_{\rm F}^2\right)
  + \mbox{h.t.} \, ,
\label{eq:TFF-convolution}
\end{eqnarray}
where $\mu_{\rm F}$ is the factorization scale between short-distance
and large-distance dynamics and h.t. denotes higher-twist
contributions.
The hard-scattering amplitude $T$ has a power-series expansion in terms
of the strong coupling
$a_s\equiv \alpha_{s}(\mu_{\rm R}^2)/4\pi$,
where $\mu_\text{R}$ is the renormalization scale.
In order to avoid scheme-dependent numerical coefficients, we set
$\mu_\text{F}=\mu_\text{R}\equiv \mu$
(default choice).\footnote{The scheme dependence of the TFF and its
sensitivity to the choice of the factorization/renormalization scale
setting was investigated in \cite{Stefanis:1998dg,Stefanis:2000vd}.}
Hence we have
\begin{equation}
  T\left(Q^2,q^2;\mu^2;x\right)
=
  T_\text{LO} + a_s~T_\text{NLO} + a_{s}^2~T_\text{NNLO} + \ldots \, ,
\label{eq:hard-scat-ampl}
\end{equation}
where the short-distance coefficients on the right-hand side can be
computed within FOPT in terms of Feynman diagrams as those depicted
in Fig.\ \ref{fig:2-photon-process}.
In our present calculation we include the following contributions,
cast in convolution form via (\ref{eq:TFF-convolution}),
and denoted by the symbol $\otimes\equiv \int_{0}^{1} dx$,
where the convenient abbreviation
$
 L
\equiv
 \ln\left[\left(Q^2y+q^2\bar{y}\right)/\mu^2\right]
$ is used \cite{Mikhailov:2009kf,Mikhailov:2016klg},
\begin{subequations}
\label{eq:T}
\begin{eqnarray}
\!  T_{\rm LO}
& \! = \! &
  T_0,
\\
\!  T_{\rm NLO}
& \! = \! & C_{\rm F}~
  T_0 \otimes \left[\mathcal{T}^{(1)}+
                      L~  V_{+}^{(0)}
              \right],
\label{eq:NLO}
\\
\!  T_{\rm NNLO}
& \! = \! &
  C_{\rm F} T_0 \otimes \!
    \left[\beta_0 T_\beta
  + T_{\Delta V}
  + T_L+  \mathcal{T}^{(2)}_c \right]\, .
\label{eq:hard-scat-series}
\end{eqnarray}
\end{subequations}
The dominant term is \cite{Melic:2002ij,Mikhailov:2009kf}
\begin{eqnarray}
  T_\beta
\!& = &\!
  \left[\mathcal{T}_\beta^{(2)} + L\left( V_{\beta +}^{(1)} -
                                      \mathcal{T}^{(1)} \right)
                                   - \frac{L^2}{2} V^{(0)}_{+}
 \right] \, ,
\label{eq:hard-scat-nnlo-beta}
\end{eqnarray}
where
$ \beta_0
=
  \frac{11}{3}{\rm C_A} - \frac{4}3 T_{\rm R} N_f
$
is the first coefficient of the QCD $\beta$ function with
$T_{\rm R}=1/2, {\rm C_F}=4/3, {\rm C_A}=3$ for $SU(3)_c$ and
$N_f$ is the number of active flavors.

Two more contributions to the NNLO radiative corrections have been
recently calculated in \cite{Mikhailov:2016klg} to which we refer for
their explicit expressions and further explanations.
These are
\begin{subequations}
 \label{eq:T-elements}
\begin{eqnarray}
   T_{\Delta V}
& \!\! = \!\! &
 L \Delta V^{(1)}_+ \, ,
 ~~~ \frac{V^{(1)}}{C_{\rm F}}
=
     \beta_0 V_{\beta}^{(1)} + \Delta V^{(1)}
\label{eq:hard-scat.nnlo-dv} \\
  T_L
& \!\! = \!\! &
 C_{\rm F} L \left[
           \frac{L}{2} V_{+}^{(0)}\otimes V_{+}^{(0)}
+ \mathcal{T}^{(1)}\otimes V_{+}^{(0)}
             \right]
                     \, ,
\label{eq:hard-scat-nnlo}
\end{eqnarray}
\end{subequations}
while the term $\mathcal{T}_{c}^{(2)}$ in (\ref{eq:hard-scat-series})
has not been computed yet and is considered in this work as the main
source of theoretical uncertainties.
Finally, suffices to say that
$V_{+}^{(0)}$ and $V_{+}^{(1)}$
are the one- and two-loop ERBL evolution kernels, whereas
$V_{\beta +}^{(1)}$
is the $\beta_0$ part of the two-loop ERBL kernel,
with
$\mathcal{T}^{(1)}$
and
$\mathcal{T}_{\beta}^{(2)}$
denoting the one-loop and two-loop $\beta_0$ parts of the
hard-scattering amplitude, respectively.

The TFF for one highly virtual photon with the hard virtuality $Q^2$
and one photon with a small virtuality $q^2\ll Q^2$ can be expressed
within the LCSR approach in the form of a dispersion integral in the
variable $q^2 \rightarrow -s$, while $Q^2$ is kept fixed, to obtain
\begin{equation}
  F_\text{LCSR}^{\gamma^*\gamma^*\pi^0}\left(Q^2,q^2\right)
=
  N_\text{T} 
  \int_{0}^{\infty}ds \frac{\rho(Q^2,s)}{q^2 + s} \, ,
\label{eq:TFF-disp-rel}
\end{equation}
where $\rho(Q^2,s)$ is the spectral density
\begin{equation}
  \rho(Q^2,s)
=
    \rho^\text{h}(Q^2,s)\theta(s_0-s)
  + \rho^\text{pert}(Q^2,s)\theta(s-s_0) \, .
\label{eq:rho-phen}
\end{equation}
The first term $\rho^\text{h}(Q^2,s)$ models the hadronic (h) content
of the spectral density,
\begin{equation}
  \rho^\text{h}(Q^2,s)
=
  \sqrt{2}f_\rho F^{\gamma^*\rho\pi}(Q^2)\delta(s-m_{\rho}^2) \, ,
\label{eq:hadr-spec-dens}
\end{equation}
while $\rho^\text{pert}(Q^2,s)$ denotes the
QCD part in terms of quarks and gluons,
calculable within perturbative QCD,
\begin{eqnarray}
  \rho^{\text{pert}}(Q^{2},s)
& = &
  \frac{1}{\pi} \text{Im} F_\text{QCD}^{\gamma^*\gamma^*\pi^0}(Q^2, -s, -i\epsilon )
\nonumber \\
& = &
   \rho_{\text{tw-2}}
  +\rho_{\text{tw-4}}
  +\rho_{\text{tw-6}}
  +\ldots\, .
\label{eq:rho-twists}
\end{eqnarray}
Each of these terms can be computed from the convolution of the
associated hard part with the corresponding DA of the same twist
(tw for short) \cite{Khodjamirian:1997tk}.
Below some effective hadronic threshold
in the vector-meson channel,
the photon emitted at large distances is replaced in
$F^{\gamma^{*}V\pi^0}$ by a vector meson $V=\rho$, $\omega$, etc.,
using for the corresponding spectral density a phenomenological
ansatz, for instance, a $\delta$-function model.

Thus, after performing the Borel transformation
$1/(s+q^2)\rightarrow \exp\left(-s/M^2\right)$,
with $M^2$ being the Borel parameter,
one obtains the following LCSR
(see \cite{Agaev:2010aq,Mikhailov:2009kf,Mikhailov:2016klg} for
more detailed expositions)
\begin{widetext}
\begin{eqnarray}
&&  Q^2 F_\text{LCSR}^{\gamma*\gamma*\pi^0}\left(Q^2,q^2\right)
=
  N_\text{T}f_\pi
  \left[
        \frac{Q^2}{m_{\rho}^2+q^2}
        \int_{x_{0}}^{1}
        \exp\left(
                  \frac{m_{\rho}^2-Q^2\bar{x}/x}{M^2}
            \right)
  \bar{\rho}(Q^2,x)
  \frac{dx}{x}
  + \! \int_{0}^{x_0} \bar{\rho}(Q^2,x)
        \frac{Q^2dx}{\bar{x}Q^2+xq^2}
  \right] \, ,
\label{eq:LCSR-FQq}
\end{eqnarray}
\end{widetext}
where the spectral density is given by
\begin{equation}
  \bar{\rho}(Q^2,s)
=
  (Q^2+s) \rho^{\text{pert}}(Q^2,s) \, .
\label{eq:rho-bar}
\end{equation}
For simplicity, we have shown the LCSR expression above using the simple
$\delta$-function model to include the $\rho$-meson resonance into
the spectral density.
However, the actual calculation of the TFF predictions to be
presented below, employs a more realistic Breit-Wigner form, as
suggested in \cite{Khodjamirian:1997tk} and used in
\cite{Mikhailov:2009kf},
\begin{equation}
  \delta(s-m_\text{V}^2)
\longrightarrow
  \Delta_\text{V}(s)
\equiv
  \frac{1}{\pi}
  \frac{m_\text{V} \Gamma_\text{V}}{(m_\text{V}^2 - s)^2
  + m_\text{V}^2 \Gamma_\text{V}^2} \, ,
\label{eq:Breit-Wigner}
\end{equation}
where the masses and widths of the $\rho$ and $\omega$ vector mesons
are given by
$m_\rho=0.770$~GeV,
$m_\omega=0.7826$~GeV,
$\Gamma_\rho=0.1502$~GeV, and
$\Gamma_\omega=0.00844$~GeV, respectively.
The other parameters entering (\ref{eq:LCSR-FQq}) are
$s =\bar{x}Q^2/x$ with $\bar{x}\equiv 1-x$,
$x_0 = Q^2/\left(Q^2+s_0\right)$, and the effective threshold in the
vector channel $s_0\simeq 1.5$~GeV$^2$.
The stability of the LCSR is ensured for values of the Borel parameter
$M^2$ in the interval $M^2\in [0.7-1.0]$~GeV$^2$
\cite{Bakulev:2011rp,Bakulev:2012nh,Stefanis:2012yw,Mikhailov:2016klg}.
By allowing a stronger variation towards larger values
$M^2\in[0.7 - 1.5]$~GeV$^2$
\cite{Agaev:2010aq,Agaev:2012tm}, the TFF prediction receives an
uncertainty of the order $[-1.6 - 7.2]\%$ \cite{Mikhailov:2016klg}.
Note at this point that the LCSR in (\ref{eq:LCSR-FQq}) includes in an
effective way the nonperturbative long-distance properties of the
real photon in terms of the duality interval $s_0$ and the
masses of the vector mesons that are absent in the pQCD formulation
of the TFF, but play an important role in the kinematic region
$Q^2\lesssim s_0$ and $x_0\lesssim 0.5$ (cf.\ the first term in Eq.\
(\ref{eq:LCSR-FQq})).

One appreciates that the real-photon limit $q^2 \to 0$ can be taken
in (\ref{eq:LCSR-FQq}) by simple substitution because there are no
massless resonances in the vector-meson channel.
Thus, this equation correctly reproduces the behavior of the TFF for a
highly virtual and a quasireal photon from the asymptotic limit
$Q^2\rightarrow \infty$ down to the hadronic normalization scale of
$Q^2\sim 1$~GeV$^2$.

Using the conformal expansion for $\rho_\text{tw-2}$, the spectral
density can be expressed in the form
\begin{eqnarray}
  \bar{\rho}\left(Q^2,x\right)
= && \!\!\!
  \sum_{n=0,2,4,\ldots}a_{n}\left(Q^2\right)
    \bar{\rho}_{n}\left(Q^2,x\right)
  + \bar{\rho}_\text{tw-4}\left(Q^2,x\right)
\nonumber \\
&& + \bar{\rho}_\text{tw-6}\left(Q^2,x\right)
+ \ldots \, ,
\label{eq:rho-bar-15}
\end{eqnarray}
where
\begin{eqnarray}
  \bar{\rho}_{n}\left(Q^2,x\right)
\!\!\!&\!\!=\!\! &\!\!\!
    \bar{\rho}_{n}^{(0)}(x)
\!\! +\!a_{s}\bar{\rho}_{n}^{(1)}(Q^2,x)
\!\!+\!a_{s}^{2}\bar{\rho}_{n}^{(2)}(Q^2,x)
    + \ldots, \nonumber \\
\bar{\rho}_{n}^{(0)}(x)&=& \psi_n(x);~\, a_s= a_{s}(Q^2) \, ,
\label{eq:rho-n}
\end{eqnarray}
with the elements $\bar{\rho}_{n}^{(i)}$ being given in Appendix B
of Ref.\ \cite{Mikhailov:2016klg}.

The dispersive analysis here includes the twist-four and
twist-six spectral densities in explicit form, as done in
\cite{Mikhailov:2016klg}.
The $\bar{\rho}_\text{tw-4}$ spectral density is given by
\begin{equation}
  \bar{\rho}_{\text{tw-4}}(Q^2,x)
=
  \frac{\delta^2_\text{tw-4}(Q^2)}{Q^2}
  x\frac{d}{dx}\varphi^{(4)}(x)\Bigg|_{x=Q^2/(Q^2+s)}\, ,
\label{eq:rho-tw-4}
\end{equation}
where the twist-four coupling parameter takes values in the range
$
 \delta^2_\text{tw-4}(\mu^{2}=1~\rm{GeV}^2)
\approx
 \lambda_{q}^{2}/2
=
 0.19\pm 0.04$~GeV$^2
$
and is closely related to the average virtuality $\lambda_{q}^{2}$
of vacuum quarks
\cite{Mikhailov:1986be,Mikhailov:1988nz,Mikhailiov:1989mk,Bakulev:1991ps,Mikhailov:1991pt},
defined by
$
 \lambda_{q}^{2}
\equiv
 \langle
        \bar{q}(ig\sigma_{\mu\nu}G^{\mu\nu})q
 \rangle/(2\langle \bar{q}q\rangle)
=
 0.4 \pm 0.05~\text{GeV}^2
$.
Details on its estimation and evolution can be found in \cite{Bakulev:2002uc},
whereas the sensitivity of the TFF to its variation was examined
in \cite{Bakulev:2003cs}.
In the present analysis the evolution of $\delta^2_\text{tw-4}$
is also included.
Expression (\ref{eq:rho-tw-4}) is evaluated with the asymptotic form of
the twist-four pion DA \cite{Khodjamirian:1997tk}
\begin{equation}
  \varphi_{\pi}^{(4)}(x,\mu^2)
= \frac{80}{3} \delta^2_\text{tw-4}(\mu^2) x^2(1-x)^2 \, ,
\label{eq:tw-4-DA}
\end{equation}
while more complicated forms were considered in
\cite{Agaev:2005rc,Bakulev:2005cp}.
The twist-six part of the spectral density, i.e.,
$\bar{\rho}_{\text{tw-6}}(Q^{2},x)
=
 (Q^2+s)\rho_{\text{tw-6}}(Q^2,s)
$,
was first derived in \cite{Agaev:2010aq}.
An independent term-by-term calculation in \cite{Mikhailov:2016klg}
provided the same result, which we quote here in the form
\begin{widetext}
\begin{eqnarray}
    \bar{\rho}_\text{tw-6}(Q^{2}\!,x)
=
    8\pi \frac{C_\text{F}}{N_c}
    \frac{ \alpha_s\langle\bar{q} q\rangle^2}{f_\pi^2}\frac{x}{Q^4}
    \left[
        \!-\!
        \left[\frac{1}{1-x}\right]_+
        \!+\!\left(2\delta(\bar{x})-4 x\right)\!+\!
        \left(
         3x+2x\log{x}
        \!+\!
        2x\log{}\bar{x}
        \right)
    \right] \, ,
\label{eq:tw-6}
\end{eqnarray}
\end{widetext}
where the plus prescription
$
 [f(x,y)]_{+}
 =
 f(x,y) - \delta(x-y)\int_{0}^{1}f(z,x) dz
$
is involved,
while
$\alpha_s=0.5$
and
$
  \langle \bar{q} q\rangle^2
=
 \left(0.242 \pm 0.01 \right)^6$ GeV$^6$ \cite{Gelhausen:2013wia}.

To obtain detailed numerical results for the TFF
$\mathcal{F}(Q^2)$
using (\ref{eq:LCSR-FQq}),
we employ several DAs from different approaches with
various shapes encoded in their conformal coefficients $a_n$.
The latter are determined at their native normalization scale
(as quoted in the referenced approaches)
by means of the moments of the pion DA
\begin{equation}
  \langle \xi^{N} \rangle_{\pi}
\equiv
  \int_{0}^{1} \varphi_{\pi}^{(2)}(x,\mu^2) (x-\bar{x})^{N}dx \, ,
\label{eq:moments}
\end{equation}
where $\xi = x - \bar{x}$ and $N=2,4, \ldots$.
The expansion coefficients $a_n$ can be expressed in terms of the
moments
$\langle\xi^N\rangle_\pi$
as follows
\begin{widetext}
\begin{eqnarray}
  a_{2n}
=
  \frac{2}{3}
  \frac{4n+3}{(2n+1)(2n+2)2^{2n}}
  \sum_{m=0}^{n} (-1)^{(n-m)}
  \frac{\Gamma(2n+2m+2)}{\Gamma(n+m+1)\Gamma(n-m+1)\Gamma(2m+1)}
  \langle \xi^{2m} \rangle_{\pi} ~~~~~ (n=0,1,2,3 \ldots) \, .
\label{eq:a-n-vs-xi-n}
\end{eqnarray}
\end{widetext}

\begin{center}
\begin{table*}[]
\caption{Conformal coefficients $a_2$, $a_4$, $a_6$ for various pion
DAs discussed in the text at two typical hadronic momentum scales
$\mu_1 = 1$~GeV and
$\mu_2=2$~GeV.
If $\mu_2$ is not the initial scale, NLO ERBL evolution in the global
scheme is employed as explained in the text and in
Appendix \ref{sec:global-NLO-evolution}.
The errors of the BMS DA given below are related to the determination
of $a_2$ and $a_4$ from QCD sum rules with nonlocal condensates.
They cause the variation of the TFF predictions shown in the form of
a green shaded band in
Figs.\ \ref{fig:linear-scaled-TFF} and
\ref{fig:omega-scaled-TFF}.
Note that the coefficient $a_2$ of the CZ DA was originally given at
the scale $\mu=0.5$~GeV: $a_{2}^\text{CZ}=2/3$ \cite{Chernyak:1983ej}.
The detailed procedure how to evaluate it at higher scales
is described in Appendix B of Ref.\ \cite{Bakulev:2002uc}.
Higher conformal coefficients up to and including $a_{12}$ for the
DSE-DB and DSE-RL DAs at the scale $\mu_2$ can be found in
\cite{Raya:2015gva}.
For the holographic AdS/QCD DA
$\varphi_{\pi}^\text{hol}(x)=(8/\pi)\sqrt{x\bar{x}}$
the coefficients up to and including $a_{20}$ at the
scale $\mu_1$ are tabulated in \cite{Brodsky:2011yv}.
They are calculable by means of the expression
$
 \left\langle\xi^{2n}\right\rangle_{\pi}^\text{AdS/QCD}
=
 \frac{1}{4}\frac{B\left(3/2,(2n+1)/2\right)}{B(3/2,3/2)}
$
[$B(x,y)$ being the Euler Beta function]
in combination with Eq.\ (\ref{eq:a-n-vs-xi-n}).
The symbol (?) in the lattice result of \cite{Braun:2015axa}
indicates that it was not extrapolated to the continuum limit.
The lattice results of \cite{Bali:2019dqc} at NNLO and NLO are quoted
separately.
They were obtained from a combined extrapolation to the chiral and
continuum limit.
}
\begin{ruledtabular}
\begin{tabular}{lccccccc}
Pion DA                                      & $a_2(\mu_1)$          & $a_4(\mu_1)$         & $a_6(\mu_1)$    & $a_2(\mu_2)$    & $a_4(\mu_2)$   & $a_6(\mu_2)$      & $\langle1/x\rangle(\mu_2)$
\\\hline \hline
BMS \cite{Bakulev:2001pa, Mikhailov:2016klg} & $0.203_{-0.057}^{+0.069}$  & $-0.143_{-0.087}^{+0.094}$  & 0 & $0.149_{-0.043}^{+0.052}$ & $-0.096_{-0.058}^{+0.063}$ & 0 & $3.16^{+0.09}_{-0.09}$
\\
BMS range                                    & $[0.146, 0.272]$      & $[-0.23, -0.049]$    & 0         & $[0.11, 0.20]$            & $[-0.15, -0.03]$               & 0 & --
\\
platykurtic \cite{Stefanis:2014nla}   & $0.0812_{-0.025}^{+0.0345}$  & $-0.0191_{-0.0287}^{0.0337}$ & 0 & $0.057^{+0.024}_{-0.019}$ & $-0.013^{+0.022}_{-0.019}$     & 0
& $3.13^{+0.14}_{-0.10}$
\\
platykurtic range                     & $[0.0562, 0.1156]$           & $[-0.0478, 0.0147]$  & 0         & $[0.04, 0.08]$            & $[-0.03, 0.01]$                & 0 & --
\\
DSE-DB \cite{Chang:2013pq}                   & --                    & --                   & --        & 0.149                     & 0.076           & 0.031        & 4.6
\\
DSE-RL \cite{Chang:2013pq}                   & --                    & --                   & --        & 0.233                     & 0.112           & 0.066        & 5.5
\\
AdS/QCD \cite{Brodsky:2011yv}                & $7/48$                & $11/192$             & $5^3/2^{12}$     & 0.107              & 0.038           & 0.0183       & 4.0
\\
Light-Front QM \cite{Choi:2014ifm}           & 0.0514                & -0.0340              & -0.0261          & 0.035              & $-0.0227$       &-0.0153       & 2.99
\\
NL$\chi$ QM \cite{Nam:2006sx}                & 0.0534                & -0.0609              & -0.0260          & 0.037              & $-0.041$        &-0.015        & 3.18
\\
CZ (this work)                               & 0.56                  & 0                    & 0                & 0.412              & 0               & 0            & 4.25
\\
Lattice \cite{Braun:2015axa}                 & --                    & --                   & --               & 0.1364(154)(145)(?)& --              & --           & --
\\
Lattice (NNLO) \cite{Bali:2019dqc}           & --                    & --                   & --               &$0.099^{+17}_{-17}(11)(11)(5)$        & --           & -- & --
\\
Lattice (NLO) \cite{Bali:2019dqc}            & --                    & --                   & --               &$0.076^{+19}_{-18}(18)(12)(4)$        & --           & -- & --
\\
\end{tabular}
\end{ruledtabular}
\label{tab:pion-DAs-a6}
\end{table*}
\end{center}

\subsection{Parton-level predictions for the TFF}
\label{subsec:parton-pred}
Let us now outline the calculational procedure to obtain predictions
for the scaled TFF
$
  \mathcal{F}(Q^2)
$
within our LCSR scheme using as nonperturbative input the various
pion DAs given in Table \ref{tab:pion-DAs-a6} in terms of their
conformal coefficients $a_2, a_4, a_6$ at the scales $\mu_1=1$~GeV
and $\mu_2=2$~GeV.
If $\mu_2$ is not the original normalization scale
of a DA, NLO evolution in the global scheme
(see App.\ \ref{sec:global-NLO-evolution}) is applied.
The number of the conformal coefficients included in the TFF
calculation depends on the particular shape of the DA.
For the two types of DAs derived from QCD sum rules with nonlocal
condensates, notably, the bimodal BMS DA \cite{Bakulev:2001pa} and the
platykurtic DA \cite{Stefanis:2014nla}, a two-parametric form
involving the lowest two nontrivial coefficients $a_2$ and $a_4$ is
sufficient.\footnote{In fact, all conformal coefficients up to and
including $a_{10}$ were computed from the first ten moments
$\langle \xi^{2n}\rangle$ ($n=1, 2, \ldots , 5)$ in
\cite{Bakulev:2001pa,Stefanis:2015qha} and the coefficients
$a_{\geqslant 6}$ were found to be compatible with zero, notably,
$a_6\approx a_2/3;
\quad a_8\approx a_2/4;
\quad a_{10}\approx a_2/5$,
but bearing large uncertainties
see Fig.\ 1 in \cite{Bakulev:2004mc}.}
Note that these DAs correspond to slightly different but admissible
values of the average vacuum-quark virtuality,
$\lambda_{q}^{2}(\mu^2\approx 1~\mbox{GeV}^2)=0.4$~GeV$^2$ and
$\lambda_{q}^{2}(\mu^2\approx 1~\mbox{GeV}^2)=0.45$~GeV$^2$,
respectively; they both have suppressed endpoint regions $x=0,1$.
The motivation underlying the construction of the platykurtic DA
\cite{Stefanis:2014nla}
originates from the desire to find a DA that generically combines
the implications of the vacuum nonlocality with the dynamical mass
dressing of the confined quark due to the DCSB.
The DA with a short-tailed platykurtic profile represents the optimal
realization of this task within the space of the conformal expansion
using QCD sum rules with nonlocal condensates, see
Fig.\ 2 in \cite{Stefanis:2015qha}.

Unimodal DAs with more or less suppressed tails have been
obtained in other approaches as well.
Recent examples are the light-front (LF) quark model (QM) of
\cite{Choi:2014ifm}
and the spin-improved holographic model of
\cite{Ahmady:2016ufq,Ahmady:2017zxf}.
It is worth reminding in this context that arguments to support the
suppression of the endpoint regions of the pion DA were already given in
\cite{Stefanis:1998dg,Stefanis:2000vd,Nam:2006sx,Choi:2007yu}
in the context of quantum fluctuations of the QCD (instanton) vacuum
and the appearance of fermionic zero modes.

\begin{figure*}[t]
 \includegraphics[width=0.60\textwidth]{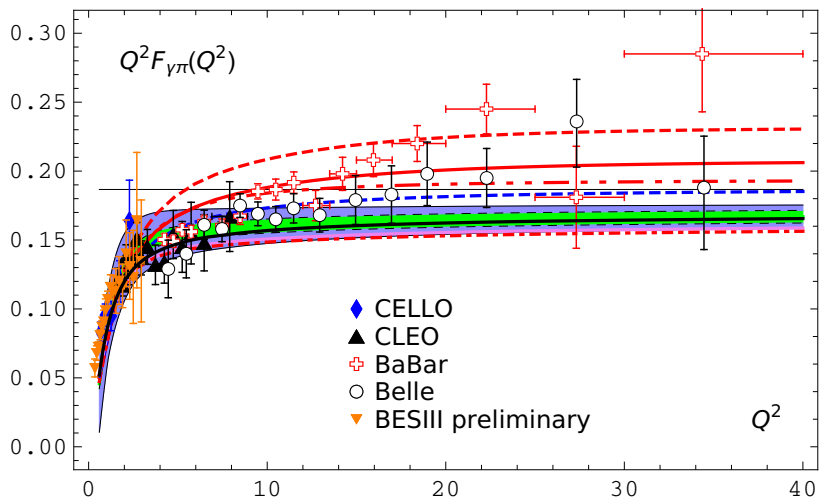} 
\caption{Measurements of the scaled pion-photon transition form
factor $Q^2F_{\gamma\pi}(Q^2)$ from different experiments, with labels
as indicated, in comparison with theoretical predictions obtained
with various pion DAs in LCSRs.
The innermost green shaded strip shows the range of predictions
pertaining to the set of the bimodal BMS DAs from
\cite{Bakulev:2001pa}, with the thick black line inside it denoting the
result for the platykurtic DA \cite{Stefanis:2014nla}.
The wider band (in blue color) around the strip encapsulates the
principal theoretical uncertainty owing to the unknown NNLO term $T_c$
(see the text for explanations).
The upper two red lines illustrate the predictions from the DSE
approach:
DSE-DB \cite{Chang:2013pq} (solid line) and DSE-RL \cite{Chang:2013pq}
(dashed line).
The dashed-dotted-dotted red line denotes the prediction obtained
for the DSE-DB DA using a lower conformal resolution
(only $\{a_2,a_4\}$).
The dashed blue line below it represents the result derived from
AdS/QCD \cite{Brodsky:2007hb}, whereas the solid pink line and the red
dashed-dotted line below the lower boundary of the total BMS band show
the predictions calculated with a light-front-based field
theoretical quark model \cite{Choi:2014ifm} and an instanton-based
chiral quark model \cite{Nam:2006sx}, respectively.
The horizontal solid line marks the asymptotic limit
$\mathcal{F}_{\infty} \approx 0.187$~GeV following from perturbative
QCD, cf.\ Eq.\ (\ref{eq:asy-TFF}).
\label{fig:linear-scaled-TFF}
}
\end{figure*}

By contrast, the calculation of the LO and NLO terms of the TFF with
broad unimodal DAs with heavy tails, requires the inclusion of all
coefficients up to $a_{12}$, see the right panel of
Fig.\ 2 in \cite{Stefanis:2014yha}, though in our approach the
contribution of $a_{12}$ is only about $0.2\%$ \cite{Stefanis:2014yha}.
Examples of such pion DAs are those derived from the
Dyson-Schwinger equations (DSE) approach of
Ref.\ \cite{Chang:2013pq} (DSE-DB and DSE-RL), where DB stands for the
most advanced Bethe-Salpeter kernel and RL denotes the rainbow ladder
approximation.
The broad shapes of these DAs are attributed to DCSB
\cite{Chang:2013pq,Raya:2015gva}.
A similarly concave pion DA with a somewhat narrower profile
was determined in the framework of holographic AdS/QCD
\cite{Brodsky:2011yv,Brodsky:2011xx}.
These three DAs together with the platykurtic one agree well with the
sum-rule estimate
$\varphi_{\pi}^\text{SR}(x=1/2)=1.2\pm 0.3$
computed in \cite{Braun:1988qv} (see \cite{Stefanis:2014yha} for the
numerical values).
A comparative illustration of the shapes of some of the mentioned pion
DAs can be found in \cite{Stefanis:2015qha}.
The corresponding graphic representations for the TFF are shown in
Fig.\ \ref{fig:linear-scaled-TFF}, where the simplified notation
$Q^2F_{\gamma\pi}(Q^2)$ is used.
These TFF predictions have been obtained by applying the NLO evolution
scheme with varying heavy flavors discussed in
App.\ \ref{sec:global-NLO-evolution}.
This scheme works for an arbitrary number of Gegenbauer coefficients,
so that such broad DAs can be evolved appropriately.
This figure is supplemented by Table \ref{tab:ff-values-table} in
App.\ \ref{sec:data-theory}, where we supply the data from the
CELLO \cite{Behrend:1990sr}, CLEO \cite{Gronberg:1997fj},
\textit{BABAR} \cite{Aubert:2009mc}, and Belle \cite{Uehara:2012ag}
experiments.
The preliminary BESIII data can be found in graphical form in
\cite{Redmer:2018uew}, see also \cite{Danilkin:2019mhd}.

We now consider our results more systematically.

(i) The range of the $\mathcal{F}(Q^2)$ values, obtained with
the set of the BMS DAs \cite{Bakulev:2001pa}, is illustrated in
Fig.\ \ref{fig:linear-scaled-TFF} by means of a green shaded band,
whereas the prediction derived with the platykurtic DA
\cite{Stefanis:2014nla} is denoted by a single solid black line.
This is because, as quantified in \cite{Stefanis:2015qha}, the margin
of variation of $a_2$ and $a_4$ for the platykurtic pion DA is much
smaller than that for the BMS DA, entailing uncertainties for the TFF
that are covered by those obtained with the BMS set of DAs.
Note, however, that the corresponding domains in the $(a_2,a_4)$
space do not overlap, see Table \ref{tab:pion-DAs-a6} and
Fig.\ 2 in \cite{Stefanis:2015qha}.
The calculation of the twist-two part of the TFF with the BMS DAs and
the platykurtic one includes the LO, NLO, and NNLO-$T_{\beta_0}$
contributions to the short-distance coefficients, cf.\ (\ref{eq:T}),
(\ref{eq:hard-scat-nnlo-beta}), at the $\{\psi_0, \psi_2, \psi_4\}$
level of the conformal expansion.
The $\psi_0$ eigenfunction yields the largest (negative)
NNLO-$T_{\beta_0}$ contribution.
Therefore, also the term $T_{\Delta V}\ll T_{\beta_0}$,
cf.\ (\ref{eq:hard-scat.nnlo-dv}),
is taken into account only via the zero harmonic $\psi_0$.
Finally, the term NNLO-$T_{L}$ vanishes for $\psi_0$,
cf.\ (\ref{eq:hard-scat-nnlo}) \cite{Mikhailov:2016klg}.
The remaining NNLO term $T_c$ is unknown and this unknownness induces
the dominant theoretical uncertainty in the TFF prediction (blue strip
enveloping the green one).
To estimate it, we assume that this term may be comparable in magnitude
to the leading NNLO term $T_{\beta_0}$
(which actually means that its potential influence is likely to be
overestimated).
This main theoretical uncertainty, computed with the BMS DAs via the
coefficients $a_2$ and $a_4$ at each measured value of the TFF, is
included in Table \ref{tab:ff-values-table}.
The analogous errors for the platykurtic DA are within these intervals
and have been omitted.
Estimates of further theoretical errors can be found in
\cite{Mikhailov:2016klg}, whereas the influence on the TFF predictions
of the virtuality of the untagged photon has been determined within our
LCSR approach in \cite{Stefanis:2012yw}.
This effect yields suppression at all momentum scales and diminishes
at high $Q^2$ so that it has little influence on our considerations
regarding the long-term $Q^2\to\infty$ behavior of the TFF.
The total TFF comprises in the spectral density
(\ref{eq:rho-bar-15}) the contributions (\ref{eq:rho-tw-4})
(twist four) and (\ref{eq:tw-6}) (twist six).
The uncertainties induced by these terms were estimated in
\cite{Mikhailov:2016klg}, see also \cite{Mikhailov:2016lof}.
They are not included in the presented predictions in
Fig.\ \ref{fig:linear-scaled-TFF} because they do not affect them
qualitatively at large $Q^2$.
Also the explicit inclusion of the small coefficient $a_6$ in the
conformal expansion of the BMS DA does not modify the TFF prediction
markedly but contributes to the theoretical noise
\cite{Stefanis:2012yw}.
We do not consider this effect here.
This notwithstanding, the inclusion of a third coefficient in the
TFF calculation increases the overlap between the BMS-based prediction
(green shaded band in Fig.\ \ref{fig:linear-scaled-TFF})
and the Belle data at high $Q^2$ \cite{Stefanis:2012yw}.

(ii) The dashed red line farthest above the asymptotic limit
(horizontal black line) in
Fig.\ \ref{fig:linear-scaled-TFF} shows the $\mathcal{F}(Q^2)$
prediction for the DSE-RL DA
\cite{Chang:2013pq}, whereas the solid red line below it exposes
the result for the DSE-DB DA \cite{Chang:2013pq}.
The corresponding twist-two TFF predictions were obtained by
incorporating the LO and NLO contributions using the set
$\{a_2, a_4, \ldots, a_{12}\}$.
The values of these conformal coefficients at the initial scale $\mu_2$
are taken from \cite{Raya:2015gva}.
The NNLO term $T_{\beta_0}$ is sufficiently included by means of
the smaller set of coefficients $\{a_2, a_4, a_6\}$, given in
Table \ref{tab:pion-DAs-a6}.
This restricted treatment already embodies over 96 percent of the
total TFF, see Fig.\ 2 (right panel) in \cite{Stefanis:2014yha}.
The other two NNLO terms, $T_{\Delta V}$ and $T_{L}$, are
treated as in item (i).
The contributions of the partial NNLO terms for the various
Gegenbauer eigenfunctions $\psi_n(x)$ have been examined in
\cite{Mikhailov:2016klg}---see Fig.\ 2 and Appendix B there.
The long-dashed-dotted-dotted red line shows the prediction for the
TFF involving the DSE-DB DA with a lower conformal resolution that
relies only upon the coefficients $a_2$ and $a_4$.
As one sees, this prediction agrees better with the asymptotic limit,
despite the fact that the reduced set of conformal coefficients
represents a rather crude approximation of the broad shape of the
DSE-DB DA \cite{Raya:2015gva} given by
$
 \varphi_{\pi}^\text{DSE-DB}(x)
=
 N_{\alpha}^\text{DB} [x\bar{x}]^{\alpha_{-}^\text{DB}}
 [1+a_{2}^\text{DB}C_{2}^{\alpha_{-}^\text{DB}+1/2}(x-\bar{x})]
$,
where
$N_{\alpha}^\text{DB}=1.81$,
$\alpha_{-}^\text{DB}=0.31$,
$a_{2}^\text{DB}=-0.12$.
Thus, the inclusion of a large number of conformal coefficients in
unimodal DAs with enhanced endpoint regions may have a detrimental
effect on the quality of the TFF prediction inside our LCSR-based
scheme.

(iii) The computation of the TFF for the holographic DA
\cite{Brodsky:2011yv} proceeds along the lines described in the
previous item.
The twist-two part includes at the LO+NLO level all coefficients
$\{a_2,\ldots,a_{12}\}$, whereas at the NNLO, the $T_{\beta_0}$
contribution comprises $(a_2,a_4,a_6)$, the term $T_{\Delta V}$
includes only $a_0$, whereas $T_{L}=0$ and $T_{c}$ is unknown,
contributing an unestimated theoretical error.
The result for $\mathcal{F}(Q^2)$ is displayed in
Fig.\ \ref{fig:linear-scaled-TFF} in terms
of a dashed blue line running close to the asymptotic limit above
10~GeV$^2$ and approaching it fast from below.
The first three coefficients $\{a_2, a_4, a_6\}$ of this DA are given
in Table \ref{tab:pion-DAs-a6} at both scales $\mu_1$ and $\mu_2$ using
NLO ERBL evolution in the global scheme to connect them.
It is worth noting that there is a misprint in Table II of
\cite{Brodsky:2011yv}.
The coefficients starting with order 10 up to 20 are actually one order
lower, i.e., 8 to 18.
For the convenience of the reader, we supply here the missing
coefficient at $\mu_1$:
$a_{20}(\mu_{1}=1~\text{GeV})= 0.0037$.

(iv) Figure \ref{fig:linear-scaled-TFF} includes the TFF
derived within our approach with the light-front quark model (LFQM)
from \cite{Choi:2014ifm} in terms of a solid line in pink color
below the blue strip.
The calculation takes into account at the twist two level the LO and
NLO contributions, as well as the radiative correction induced by the
term NNLO-$T_{\beta_0}$ using as a nonperturbative input the conformal
coefficients $\{a_2, a_4, a_6\}$, computed by the authors at the
initial scale $\mu_1$ (see Table II in \cite{Choi:2014ifm}).
The numerical values of these coefficients at $\mu_2$, shown in Table
\ref{tab:pion-DAs-a6} and used in our graphics, were obtained with
NLO ERBL evolution in the global scheme.
The other NNLO term, notably, $T_{\Delta V}$, is included for the
zero harmonic, whereas, as before, $T_{L}=0$, and the error induced by
the unknown contribution $T_c$ is ignored.

(v) Table \ref{tab:pion-DAs-a6} also contains the conformal
coefficients $\{a_2, a_4, a_6\}$ of the nonlocal chiral quark model
(NL$\chi$QM) from \cite{Nam:2006sx} (model 3 in their Table II) at the
two scales $\mu_1$ and $\mu_2$ (the latter after NLO evolution in the
global scheme).
The intrinsic scale of this model is actually lower than $\mu_1$, but
we follow the line of reasoning given by the authors to show the
conformal coefficients at the normalization scale $\mu_1$
(dubbed $\Lambda_1$ in their terminology) with their original values
\cite{Nam:2006sx}.
One observes from Table \ref{tab:pion-DAs-a6} that these coefficients
are quite close to those of the light-front quark model
\cite{Choi:2014ifm} treated in item (iv).
As a result, the obtained TFF prediction
(dashed-dotted red line below all other curves) in
Fig.\ \ref{fig:linear-scaled-TFF}
practically coincides with the LFQM one
(solid pink line just above it).

(vi) We shortly comment here on the recent lattice results
for the second Gegenbauer moment from \cite{Bali:2019dqc}.
The quoted values were obtained from a combined chiral and
continuum limit extrapolation at the NNLO (two-loops)
and NLO (one-loop).
The corresponding central values deviate considerably from each other.
This treatment differs from that applied in \cite{Braun:2015axa},
where no extrapolation to the continuum limit was carried out.
This is indicated by the label (?) in Table \ref{tab:pion-DAs-a6}.
From this table one observes that the following DAs
have $a_2(\mu_2)$ coefficients coming close to the NNLO value
(deviation in parenthesis):
AdS/QCD (0.008), platykurtic (-0.042),
NL$\chi$QM (-0.062), and LFQM (-0.064).
As we will see later in terms of
Figs.\ \ref{fig:linear-scaled-TFF}, \ref{fig:omega-scaled-TFF}
more explicitly, the first of these DAs yields a TFF prediction that
approaches fast $\mathcal{F}_\infty$ and, depending on the number of
the conformal coefficients included in the calculation, eventually
crosses it around 60~GeV$^2$.
The other three DA models lead to predictions that remain below this
limit, with the platykurtic DA giving a TFF closest to
$\mathcal{F}_\infty$ and approaching it uniformly from below.
We mention without further discussion that the new lattice NNLO result
agrees with the value found in \cite{Agaev:2012tm} by fitting model II
\cite{Agaev:2010aq} to the Belle data instead of \textit{BABAR}.
This model gives $a_2=a_4=a_6=0.10$, while $a_8=0.034$.
It is worth noting that the previous lattice estimate
\cite{Braun:2015axa}
was quite larger (0.136) and close to the value $a_2(\mu_2)=0.140$
for the mentioned model II, but also not far away from the DSE-DB
and the BMS DA, which both give 0.149 (Table \ref{tab:pion-DAs-a6}).
Within the reported uncertainties, all these DAs are reasonable
candidates and cannot be differentiated by this single constraint.

\subsection{Asymptotic behavior of the calculated TFF}
\label{subsec:asy-TFF}
Going further, we now give a broader discussion of the
various TFF predictions, founded upon the main observations from
Fig.\ \ref{fig:linear-scaled-TFF}, focusing our attention on their
asymptotic behavior.

a) Pion DAs, like the BMS set \cite{Bakulev:2001pa}
(narrower green strip)
and the platykurtic one \cite{Stefanis:2014nla}
(solid black line inside it),
which implicitly incorporate a nonvanishing average virtuality of the
vacuum quarks, have suppressed tails at $x=0,1$ --- irrespective of
their topology at the central point $x=1/2$
(bimodal the first, unimodal the second).
This suppression entails a balanced magnitude of the scaled TFF so that
it approaches with increasing $Q^2$ the QCD asymptotic limit
monotonically from below, being at the same time in good overall
agreement within the margin of experimental and theoretical error
with those data that support QCD scaling at large $Q^2$.
More specifically, the predictions for $\mathcal{F}(Q^2)$ herewith
follow the steep rise indicated by all existing data below 4~GeV$^2$
and start to scale with $Q^2$ above about 8~GeV$^2$.
The magnitude of the scaled TFF tends toward
$\mathcal{F}_\infty$
very slowly from below.
Both types of DAs comply with the \textit{BABAR} data below 10~GeV$^2$
but deviate from them significantly above this mark (except at
$Q^2=27.31$~GeV$^2$ where they agree).
They are also overall consistent with the Belle data --- except at
$Q^2=34.36$~GeV$^2$ where they fall short.
In aggregate, the BMS band of the TFF predictions can
accommodate within the margin of experimental error all data below
15~GeV$^2$.
Note that the TFF prediction obtained with the Chernyak-Zhitnitsky
pion DA (not shown in Fig.\ \ref{fig:linear-scaled-TFF}) exceeds all
data considerably already in the region of a few GeV$^2$ and scales
then towards higher $Q^2$ values, crossing the \textit{BABAR}
branch at 22.28~GeV$^2$ and the Belle measurement at 27.33~GeV$^2$
(see \cite{Bakulev:2012nh} for a graphic illustration and
\cite{Wang:2017ijn} for a different conclusion using another
approach).

b) As we already mentioned, the LFQM-based DA
\cite{Choi:2014ifm,Choi:2017zxn} yields a TFF prediction
(solid pink line) that exhibits a
$Q^2$ behavior similar to that obtained with the BMS/platykurtic DA,
albeit with a somewhat smaller magnitude.
Also the NL$\chi$QM DA \cite{Nam:2006sx} yields a TFF
(dashed-dotted red line) which replicates this result.
The root for this proximity of predictions can be traced to the similar
profiles of the underlying pion DAs.
However, there is a crucial difference.
While the platykurtic DA has $a_6=0$, the other two models involve
a negative $a_6$ coefficient (see Table \ref{tab:pion-DAs-a6})
that causes a relative reduction of $\mathcal{F}(Q^2)$.
We remark without displaying the corresponding TFF prediction in Fig.\
\ref{fig:linear-scaled-TFF} that the spin-improved holographic DA from
\cite{Ahmady:2016ufq} yields a result which overlaps with the BMS
strip, while the underlying DA has a profile bearing similarities to
the platykurtic one.
In this sense, the term \emph{platykurtic pion DA} denotes not only
the particular DA determined in \cite{Stefanis:2014nla}, but it
connotes all unimodal DAs with suppressed tails.
However, the fine details of the TFF predictions may somewhat differ
from each other, although the variation is limited below
the asymptotic limit, see Fig.\ \ref{fig:linear-scaled-TFF}.
Note parenthetically, that a simultaneous fit \cite{Zhong:2015nxa}
to the CLEO \cite{CLEO98} and Belle \cite{Belle12} data gives
rise to a pion DA that closely resembles the platykurtic one.

\begin{figure*}[t]
\includegraphics[width=0.60\textwidth]{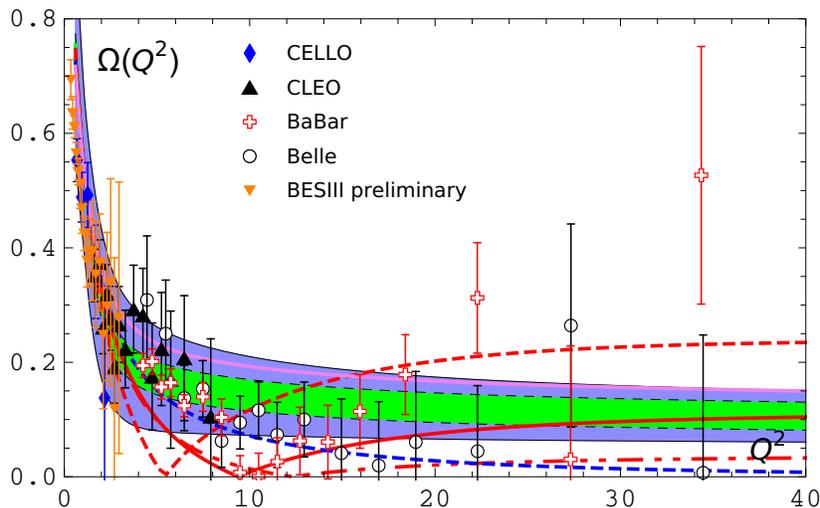}
\caption{Plot of the scaling-rate quantity $\Omega(Q^2)$,
cf.\ Eq.\ (\ref{eq:scaling-rate}),
in the $Q^2$ range $[0, 40]$~GeV$^2$ for various theoretical
predictions in comparison with the data.
The same designations as in Fig.\ \ref{fig:linear-scaled-TFF}
are used.
Further comments are given in the text.
\label{fig:omega-scaled-TFF}
}
\end{figure*}

c) The TFF predictions obtained in our scheme with broad unimodal DAs
with heavy tails can vary considerably.
The prediction based on the holographic AdS/QCD DA
\cite{Brodsky:2011yv} appears to be closest to the BMS strip and
provides a rather good agreement with the Belle data, while it
disagrees with the \textit{BABAR} data above
10~GeV$^2$, where these start to grow.
It approaches $\mathcal{F}_\infty$ from below already near 40~GeV$^2$.
In this sense, the AdS/QCD DA represents a crossover form from the
platykurtic DA to still broader unimodal DAs with heavy tails.
A follow-up extension of the AdS/QCD approach \cite{Chang:2016ouf}
employs smaller effective quark masses to handle endpoint singularities
and proposes a platykurtic-like DA structure and TFF prediction
(Scenario 1),
whereas another alternative
(Scenario 2)
amounts to a broad DA with heavy tails yielding a TFF prediction
that crosses the asymptotic limit already just above 15~GeV$^2$.
Note that the mentioned spin-improved version of the AdS/QCD DA in
Ref.\ \cite{Ahmady:2016ufq} receives less endpoint enhancement so that
the associated TFF prediction approaches $\mathcal{F}_\infty$ from
below without crossing it.

d) Staying within the context of broad unimodal DAs, let us mention
that the results for the TFF computed with the DSE DAs within our
LCSR scheme show a strong sensitivity to the power $\alpha_{-}$
in the ``Gegenbauer-$\alpha$'' representation which employs Gegenbauer
polynomials of variable dimensionality $\alpha=\alpha_{-}+1/2$
\cite{Chang:2013pq,Gao:2014bca}
\begin{equation}
  \varphi_{\pi}^{(\alpha)}(x,\mu^2)
=
  N_\alpha (x\bar{x})^{\alpha_{-}}
  [1 + a_{2}^{\alpha}C_{2}^{(\alpha)}(x-\bar{x})] \, .
\label{eq:DSE-DA}
\end{equation}
The TFF predictions obtained with these DAs have magnitudes
that grow in inverse proportion to $\alpha_{-}$.
Indeed, varying $\alpha_{-}$ from the value $0.5$ (AdS/QCD) to
$0.31$ (DSE-DB) to $0.29$ (DSE-RL) causes, say, at $Q^2=30$~GeV$^2$,
an increase of the TFF magnitude between about $10\%$ and $20\%$.

The above remarks on the DSE-based TFF predictions should be taken
with caution.
The reason is that our findings were obtained within the LCSR-based
scheme following the calculational procedure described above.
It was argued in \cite{Raya:2015gva} that calculating the TFF entirely
within the DSE framework, it is possible to relate it to the
perturbative QCD prediction for the transition form factor in the
hard-photon limit.
As a result, the DSE-DB DA-based prediction, obtained this way
in \cite{Raya:2015gva}, still belongs to the class of predictions
compatible with scaling, see Fig.\ 2 in \cite{Bakulev:2012nh}.

To explore more profoundly the long-term behavior of the TFF and its
scaling rate at high $Q^2$, it is convenient to define the following
quantity
\begin{equation}
  \Omega(Q^2)
\equiv
  \frac{\left|\mathcal{F}(Q^2) - \mathcal{F}_{\infty}\right|}%
  {\mathcal{F}_{\infty}}
\label{eq:scaling-rate}
\end{equation}
that provides a normalized measure of the deviation of the scaled
form factor from the asymptotic value
$\mathcal{F}_{\infty}=\sqrt{2}f_\pi$---the ``baseline''.
The graphical representation of the theoretical results for this
quantity versus $Q^2$ in comparison with the data is shown in
Fig.\ \ref{fig:omega-scaled-TFF}.
The designations are the same as in Fig.\ \ref{fig:linear-scaled-TFF}.

One immediately appreciates from this figure that the TFF predictions
and their principal intrinsic uncertainties obtained in our scheme with
the set of the BMS DAs
(shaded bands in green and blue color, respectively)
and the solid black line, which denotes the TFF for the platykurtic DA,
tend to the limit $\mathcal{F}_{\infty}$ uniformly and do no reach it
even at 40~GeV$^2$.
A similar scaling rate, though further above the baseline, is
exhibited by the TFF for the LFQM DA \cite{Choi:2014ifm}
(with the NLCQM DA \cite{Nam:2006sx} being intimately close to
it---therefore not shown).
The AdS/QCD-based TFF prediction \cite{Brodsky:2011yv} reaches the
asymptotic value much faster than the said curves showing a clear
tendency for complete saturation.
In the considered approximation, it crosses the baseline at 57~GeV$^2$.
The DSE-based predictions \cite{Chang:2013pq}, which include all
conformal coefficients from $a_2$ to $a_{12}$ intersect the baseline
already at $Q^2\approx 4$~GeV$^2$ (DSE-RL) and
$Q^2\approx 10$~GeV$^2$ (DSE-DB).
Because $\Omega(Q^2)$ is defined as the modulus, the corresponding TFF
curves seem to bounce off at the crossing point with the baseline.
What's more, these curves continue to deviate from $\Omega(Q^2)$
more and more as the momentum increases, though it is unclear
whether they may return to the baseline at some remote $Q^2$ value.
As to the model II from \cite{Agaev:2010aq}, it yields a prediction
(not shown in Fig.\ \ref{fig:omega-scaled-TFF}) that would bounce off
the baseline and follow a trend similar to the solid red line, while
its modified version \cite{Agaev:2012tm} would be inside the larger
band in blue color of the BMS predictions.

From the experimental side, one appreciates that several high-$Q^2$
data points of the \textit{BABAR} experiment also move away from
the baseline at a rather fast rate.
Even their error bars are far away from the baseline, so that no
saturation of the TFF can be inferred from this data set.
The behavior of this branch of the \textit{BABAR} data does not look
haphazard and erratic but---within the reported errors---rather
systematic and self-generated.
Its origin is the subject of several theoretical investigations
(see \cite{Bakulev:2012nh} for a detailed discussion and references).
In contrast, the error bars of the Belle outlier at 27.33~GeV$^2$
come quite close to the baseline, and the ultimate Belle data point
at 32.46~GeV$^2$ coincides exactly within its error margin with the
baseline, which one may interpret as an indication for saturation
of the TFF at this scale, though this agreement may be purely
coincidental.
Both data sets (\textit{BABAR} and Belle) have rather poor statistics
at high $Q^2$.

On the other hand, there is no data with high statistical
precision at the low end of $Q^2$, say, below 4~GeV$^2$.
To give an impression of the current situation in this domain,
we have included in Figs.\ \ref{fig:linear-scaled-TFF} and
\ref{fig:omega-scaled-TFF}, the preliminary data of the BESIII
collaboration \cite{Redmer:2018uew}.
They cover the momentum range $[0.3-3.1]$~GeV$^2$ that lies well below
$\mu_2^2=4$~GeV$^2$ and bear a rather large total error at the upper
end.
In contrast, they exceed the accuracy of the CELLO data towards
low-$Q^2$ values.
This low-$Q^2$ domain of the TFF is characterized by a rapid growth of
$\mathcal{F}(Q^2)$ (Fig.\ \ref{fig:linear-scaled-TFF}) or,
equivalently, a steep fall-off of $\Omega(Q^2)$ before reaching the
scaling regime represented by the baseline
(Fig.\ \ref{fig:omega-scaled-TFF}).
High-precision data in the low-momentum domain of $Q^2$, between the
scales $\mu_1$ and $\mu_2$, would be extremely useful in order to fix
the slope of the form factor with higher accuracy.
As one sees from Fig.\ \ref{fig:omega-scaled-TFF}, the predictions
obtained with the LFQM DA (pink solid line)
and the DSE-RL DA (red dashed line)
are both within the margin of error of the preliminary BESIII
data but their subsequent behavior above 4~GeV$^2$ differs dramatically.
While the LFQM DA-based prediction remains far away from the baseline
in the whole $Q^2$ domain, the DSE-RL DA leads to a TFF which crosses
the baseline below 5~GeV$^2$ and then strays away from it.
The final BESIII data at the BEPCII collider may reach a
higher accuracy by taking into account radiative effects of QED in the
efficiency corrections \cite{Danilkin:2019mhd}.
This would provide more stringent constraints on the variation of the TFF
slope in this crucial $Q^2$ region.
They could also provide clues to the contribution of additional power
corrections \cite{Wang:2017ijn,Shen:2019vdc,Shen:2019zvh}
that have not been considered in the present analysis.

\section{Phase-space reconstruction of the TFF from the data}
\label{sec:data-analysis}
Up to now we have confronted the experimental data on
$\mathcal{F}(Q^2)$
by employing methods that rely upon some theoretical basis
at the QCD level of description, e.g., \hbox{LCSRs}.
At this level, the measured process
$e^{+}e^{-}\rightarrow e^{+}e^{-}\pi^0$
involves the elementary subprocess $\gamma^*\gamma\rightarrow \pi^0$
that is described by the TFF $F^{\gamma^*\gamma\pi^0}(Q^2, q^2\to 0)$
in terms of quarks and gluons (see Fig.\ \ref{fig:2-photon-process}).
The restriction to this single variable to describe the experimental
system is justified by the fact that it can provide substantial
information on the pion DA.
Unfortunately, our considerations in the previous section show that
even the onset of the asymptotic behavior of the TFF depends in a
sensitive way on the calculational scheme involved,
cf.\ Fig.\ \ref{fig:omega-scaled-TFF}, allowing no reliable conclusion
about the shape of the pion DA.
Moreover, owing to the QCD content of $\mathcal{F}(Q^2)$,
even a best-fit procedure to reproduce the dynamical origin of the
pattern of the experimental data, obtained by particular experiments,
involves some tacit theoretical assumption underlying the adopted form
of the fit function, for instance,
a dipole formula \cite{Uehara:2012ag},
or a power function \cite{Aubert:2009mc},
which both correspond to distinctive dynamical mechanisms at the
parton level \cite{Stefanis:2012yw,Zhong:2015nxa,Czyz:2017veo}.

From the experimental side, to reveal the asymptotic behavior of the
TFF, one needs data from repeated measurements at higher and higher
values of $Q^2$ under similar conditions of high statistical quality.
Such measurements may become increasingly difficult both as regards
their technical feasibility as well as from the point of view of
their accuracy (closeness to the true value)
and precision (closeness of measurements to a single value).
But also the interpretation of data with unknown (or unsettled)
accuracy but small statistical variability are a challenge for theory
because this lack of knowledge reduces the information from experiment
to the level of observations with contextual explanations.
In fact, the high-$Q^2$ data tails of the Belle and the
\textit{BABAR} measurements show mutually incompatible trends of
the scale behavior of the form factor
(see \cite{Bakulev:2012nh,Stefanis:2012yw} for quantification).
While the Belle measurement is supporting a saturating TFF at
high-$Q^2$ (i.e., scaling), the \textit{BABAR} data points beyond
10~GeV$^2$ indicate an anomalous growth of $\mathcal{F}(Q^2)$
induced by a residual $Q^2$ dependence (of unknown origin) that
prevents agreement with the asymptotic limit of perturbative QCD.
In addition, both data sets include outliers showing at the same
momentum value of 27.3~GeV$^2$ exactly the opposite $Q^2$ behavior
relative to their respective overall trend in this region
(see Table \ref{tab:ff-values-table} and
Figs.\ \ref{fig:linear-scaled-TFF}, \ref{fig:omega-scaled-TFF})
making a proper interpretation even more difficult.
We are not going to settle the issue here.

We discuss instead a way to circumvent these problems by
analyzing the TFF data using techniques based on Takens' time-delay
embedding theorem \cite{Takens:1981}, the focus being on the
long-term (i.e., large-$Q^2$) behavior of the data.
The key idea to reveal the claimed saturating behavior of
$\mathcal{F}(Q^2)$
in the asymptotic $Q^2$ regime is to reconstruct the dynamical
evolution of the system from scalar measurements of this single
quantity using the method of delay-coordinate diffeomorphic
embedding \cite{Packard:1980zz,Takens:1981} in terms of
an attractor.\footnote{The attractor reveals (``measures'') the amount
of determinism hidden in the events which give rise to the
\emph{macroscopic} system, described by the measured quantity within
its space of states.}.
The compatibility of the results of this data analysis
with the QCD predictions obtained in Sec.\ \ref{sec:analysis-theory}
will be discussed in Sec.\ \ref{sec:pred-forecasts}.

Before delving into the details of the delay embedding
technique, it is instructive to summarize the main theoretical and
practical benefits of reconstructing the attractor.
The limitations of the method will be addressed in the context of our
analysis.

\begin{itemize}
\item
The phase-space portrait of the attractor, in terms of lagged
scatter plots, provides a \emph{global picture}
of all possible states (phases) of the considered dynamical system.

\item
Lagged phase-space plots can reveal
\emph{deterministic behavior}
in the data related to systems that have three or fewer dimensions,
even if the data in regular space appear to be randomly distributed
lacking any obvious order or pattern.

\item
The attractor represents a mathematical model for
\emph{data compression}.
This is especially important in analyzing large sets of data.

\item
The existence of an aggregation of states in the long-term
structure of the attractor in terms of lagged scatter plots
provides a diagnostic tool to reveal
\emph{saturation}
of the measured variable in this dynamical regime.
This is key in our case in unveiling the onset of the
asymptotic behavior of the TFF at much lower momenta than it is
possible
by scattered measurements at solitary high-$Q^2$ values.

\item
The reconstructed attractor can be used to make
\emph{forecasts}
about the future trend of a scalar time series of measurements.
\end{itemize}

\subsection{Topological time-series analysis and the method of delays}
\label{subsec:attr-recon-exp-data}
The temporal evolution of dynamical systems occurring in nature can be
measured and recorded in terms of a continuous or discrete time series,
i.e., a scalar sequence of measurements over time.
Each state (phase) of the dynamical system is uniquely specified by
a point or vector
$\mathbf{y}=(y_1,y_2, \ldots)$
in phase space $S$ that flows from an initial value
$\mathbf{y}_0$ to $\mathbf{y}(t)=\varphi_{t}\mathbf{y}_0$,
where $\varphi_t$ represents a one-parameter family of maps of $S$ into
itself, collectively written as $\varphi_{t}S$ \cite{Broomhead:1985}.
The evolution of the system is controlled by the
vector field $\mathbf{F}(y)$,
acting on points in $S$ for each time $t$, connecting them along
a trajectory subject to
$d\mathbf{y}(t)/dt=\mathbf{F}(y)$.
The entirety of the maps $\varphi_{t}S$
provides a global picture of the
solutions to all possible initial states of the system, although the
phase space cannot be revealed in an experiment.
The dimension of the set $\{\varphi_{t}S\}$ will initially be
the same as the dimension of $S$, which is determined by the total
(possibly unknown) number of variables describing the system.
However, it is a common phenomenon of real dynamical systems that their
evolution will cause the flow of points in $S$ to contract onto sets of
lower dimensions called phase-space attractors.\footnote{The underlying
working hypothesis is that nature is deterministic, dissipative, and
nonlinear.}
The reconstruction of the system trajectories on the attractor from
(discrete) measurements is of fundamental importance because it provides
a dynamical understanding of the time evolution of the system under
consideration that generates them.
Moreover, because the attractor exists on a
smooth submanifold $M$ of $S$
with $\text{dim}[M]<\text{dim}[S]$, the system has fewer degrees of
freedom on the attractor and consequently it requires less information
to specify its structure (phase portrait).
As a result, the attractor provides a mathematical model to reveal the
system dynamics from a compressed set of data and its visual character
facilitates its interpretation.
To achieve this goal, we have to ensure that the map from the true
(unknown) attractor in $S$ into the reconstruction space is an
embedding that preserves differentiable equivalence.

The mathematical basis for the mentioned procedure is provided by
Takens' embedding theorem \cite{Takens:1981}, which we now
explain.\footnote{The proof of this theorem is outside the scope of the
present work, see \cite{Takens:1981,Sauer:1991}.}

Let $M$ be a compact manifold of dimension $m$, $F$ a smooth ($C^2$)
vector field, and $v$ a smooth function on $M$,
$v\in C^2(M,\mathbb{R})$.
It is a generic property that
$
 \Phi_{F,v}: M \longrightarrow \mathbb{R}^{n\geq 2m+1}
$
is an embedding, where $\varphi_{t}\in C^2(M)$ is the flow of
$F$ on $M$, defined by \cite{Broomhead:1985}
\begin{equation}
  \Phi_{F,v}(y)
=
  \left[
        v(y), v(\varphi_{1}(y)), v(\varphi_{2}(y)), \ldots v(\varphi_{2m}(y))
  \right]^T \, ,
\label{eq:embedding}
\end{equation}
where $g(f(x))=(g\circ f)(x)$ is the composite function of
$f:X\longrightarrow Y$ and $g: Y\longrightarrow Z$ for all
$x$ in $X$, and $T$ denotes the transpose.
One can construct a shadow version of the original manifold $M$,
$\Phi(M)=M_v$, from a single scalar time series by shifting its
argument by an amount $\tau$ (the ``lag'') to obtain from the basic
series $v(t)$ lagged copy series, e.g.,
$v(t-\tau)$ (lagged-one series),
$v(t-2\tau)$ (lagged-two series), and so on.\footnote{Mathematically,
one can equivalently use a forward lag, but we prefer the backward-lag
notation that conforms with causality \cite{Bradley:2015nts}.}
This way, the manifold $M_v$ stores the whole history of the
measurements $v(y)$ made on the system in a state $y$
in terms of the reconstructed attractor
$
 \mathbf{X}(t)
=
 \left[v(t), v(t-\tau), v(t-2\tau), \ldots v(t-2m\tau)\right]^T
$.
The reconstruction preserves certain mathematical properties of the
original system such as the differentiable equivalence relation of
$M$ and its Ljapunov exponents
(see \cite{Sauer:1991} for more mathematical details).
More important for applications is the one-to-one mapping relation
between the original manifold $M$ and the shadow manifold $M_v$
that enables one to recover states of the original dynamics by using
a single scalar time series as an $\omega$-limit data set,
i.e., a forward trajectory (analogously, the backward trajectory is
called the $\alpha$-limit set).

In this sense, one can discern the \emph{dynamically}
generated pattern inside the \emph{geometrical} pattern of trajectories
on the attractor even if this reconstruction is not identical with the
full internal dynamics.
To unfold the intrinsic structure of the attractor, one has to estimate
the minimum embedding dimension $n$
(using, for instance, the method of the false nearest
neighbors)\footnote{These are attractor points appearing to be close to
each other with an embedding dimension $n$ but distant with the
next higher dimension.}
and determine the optimal lag $\tau$
in terms of the data autocorrelation function \cite{Bradley:2015nts}.
Metaphorically speaking, the unfolding of the attractor structure
in terms of the reconstruction (i.e., embedding) parameters $(n,\tau)$
resembles the focusing of a camera---particular combinations may
yield better results than others \cite{Huffaker:2010}.
In the present exploratory investigation, we select these embedding
parameters pragmatically without pursuing the mentioned mathematical
approaches or computer algorithms based on them \cite{Hegger:1998tis}.
We also ignore experimental errors (noise) in the attractor
reconstruction.
These refinements will be considered in future work.

We now describe the calculational steps for the attractor
reconstruction using the time series sampled from the data records
in the
CLEO \cite{Gronberg:1997fj},
\textit{BABAR} \cite{Aubert:2009mc}, and
Belle \cite{Uehara:2012ag}
experiments (Table \ref{tab:ff-values-table}).
The number of the CELLO \cite{Behrend:1990sr} data is not
sufficient to reconstruct a trajectory, while the preliminary BESIII
\cite{Redmer:2018uew} data are obtained in a very low $Q^2$ domain
that is not relevant for the asymptotic behavior of the TFF.
Each of these time series will give rise to a different attractor with
its own systematical and statistical uncertainties
(not included in the analysis).
Nevertheless, because the selected measurements were mostly performed
at the same $Q^2$ values and the involved time delays are taken to be
equal, we treat the embedding of all three time series within the same
three-dimensional (3D) phase space and plot the results together.
Mathematically speaking, our proposition is to find out wether the
probed experimental system gives rise by self-organization to system
evolution on the same low-dimensional manifolds with attractors
contained in a restricted phase-space region ( a ``corridor'')
characterized by the common embedding parameters $(n,\tau)$.
In particular, we are interested in the long-term behavior of these
attractors in order to deduce the existence of a common trajectory
regime in the vicinity of $\mathcal{F_\infty}$, where the delayed
vectors pertaining to these experiments condense on neighboring
measurements made on the system $\mathcal{F}(Q^2)$.

\begin{figure}[t]
 \includegraphics[width=0.40\textwidth]{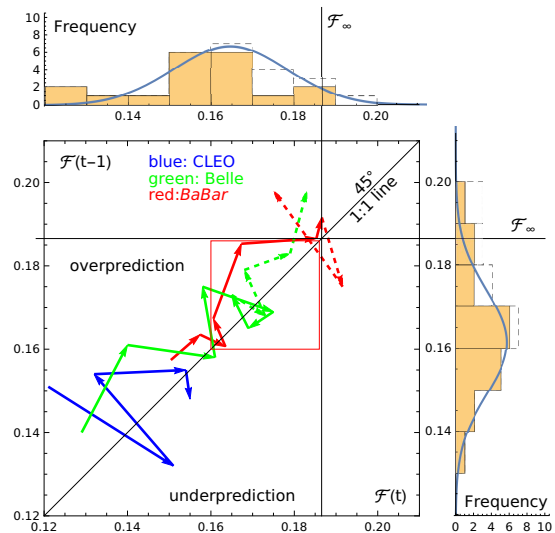}
  \caption{
\label{fig:xy-scatter-histo}
2D projection in the phase space of the reconstructed attractors
in terms of lagged coordinate vectors extracted from different data
sets using for the basic series a constant sampling length,
$\tau_s=1$~GeV$^2$.
The assignment of the attractors to the data from bottom to top
is as follows:
CLEO (blue lines) \cite{Gronberg:1997fj},
Belle (green lines) \cite{Uehara:2012ag}, and
\textit{BABAR} (red lines) \cite{Aubert:2009mc}.
The underlined time series
$\mathcal{F}(t)$ (basic series),
$\mathcal{F}(t-1)$ (lag-one series), and
$\mathcal{F}(t-2)$ (lag-two series)
are specified in the text.
The shaded histograms count the frequency of the combined vector
occurrences in each bin across the range of values common to all
three attractors.
Data values at the intersection of two bins, are placed in the next
higher bin.
The dashed lines signify additional vectors from data points
measured at somewhat larger time intervals than $\tau_s$
(see the text).
They are not included in the shaded histograms and in the
distributions, but are taken into account in the histograms bounded
by dashed lines.
The red square shows the region bounded from above by the asymptotic
TFF value $\mathcal{F}_\infty$ and adjusted to contain most vectors.
}
\end{figure}

We begin by introducing the technique in terms of a discrete sampling
of some generic experimental data set on the single dynamical variable
$\mathcal{F}(Q^2)$/GeV
(omitting the dimension in the following discussion)
and represent it as a time series of observations
$\left( y(t_1), y(t_2), y(t_3), \ldots ,y(t_{N_s}) \right)$,
where 
$N_s$ is the total number of samples in the set $s$.
To construct a basic time series $\mathcal{F}(t_i)$
we define a fixed sampling time
$\tau_s=y(t_{i+1})-y(t_{i})=1$~GeV$^2$
and select a regular subseries of elements at equidistant time scales
separated by $\tau_s=1$~GeV$^2$.
The lag $\tau$ is the time difference between the time series we are
correlating.
For convenience, we define the lag time in terms of a
parameter $J$ that counts the omitted elements within the delayed
time series, $\tau=J\tau_s$.
The basic time series corresponds to
$J=0$, the one-lagged time series
(one observation removed from the basic series) is obtained for $J=1$,
the two-lagged series
(offset from the basic series by two observations) for $J=2$ and so on,
while the total number of observations in each lagged series is $N_s-J$.
Then, the elements visible in the $(n,J)$-window \cite{Broomhead:1985}
constitute the components of a vector in the embedding space
$\mathbb{R}^n$ with the embedding dimension
$n\geq 2m+1$ \cite{Takens:1981}.
To avoid false projections by embedding the system $\mathcal{F}$
in too few dimensions, we chose in this work $n=3$ that proves to be
sufficient to unfold the structure of the attractor and enables at the
same time its geometric visualization.

The choice of the time delay $\tau$ is arbitrary, but there
are limitations, see, \cite{Pecora:2007} and references cited therein.
Adopting overly short lags, would induce strong correlations between
the time-series we compare, causing the attractor to collapse on the
$1:1$-line at 45$^\circ$ in the embedding phase space, so that the
delayed vectors cannot be resolved (almost linear regression).
On the other hand, increasing significantly the delay would average
out existing correlations between adjacent events but would eventually
contain too little mutual information to reconstruct the attractor,
making it impossible to discover any determinism inside the data
(like in a random time series).
The optimal delay choice should ensure the statistical independence
of distant time-series neighbors maintaining at the same time the right
amount of feedback to keep lagged events connected to each other.
This renders the number of the state vectors sufficiently large
in order to enable the attractor reconstruction,
avoiding a featureless thus uninteresting portrait.
Finally, if the number of events is small, like in our case, choosing
a large lag $\tau >2$ would reduce the number of state vectors
in the lagging process dramatically.
This would obscure the tell-tale characteristics of the attractor
considerably and reduce its usefulness.

Once the basic series $\mathcal{F}(t)$ for each data set has been
selected, one can create additional time series with $N-J$ elements
by displacing the time value and shifting the basic time
series for its entire length by one unit ($J=1$), two units ($J=2$),
etc.
This way, a sequence of $n$ lagged coordinate vectors in the
embedding space
$\mathbb{R}^n$ can be generated:
$\mathcal{F}(t-1)$ (lag-one~series),
$\mathcal{F}(t-2)$ (lag-two~series) and so on,
where we used the convenient notation
$\mathcal{F}(t_i-J\tau)=\mathcal{F}(t-J)$ with $\tau=1$.
The lagged vectors on the attractor form a discrete trajectory, i.e.,
a scatter plot in $\mathbb{R}^n$, given by the expression
\begin{eqnarray}
  \mathbf{X}_i
& = &
  \Phi_{\mathbf{F},\mathbf{v}}\left(\varphi_t(\mathbf{y})\right)
\nonumber \\
& = &
  \left(
  v_i , v_{i+J} , v_{i+2J} , v_{i+3J} , \ldots , v_{i+(n-1)J}
  \right)^\text{T} \, .
\label{eq:trajectory}
\end{eqnarray}

Here $\mathbf{X}_i$ can be a column vector, representing a univariate
time series with $N_s$ data points, or a trajectory matrix composed of
a multivariate time series in terms of a sequence of
$N=N_{s}-(n-1)$ vectors,
$\{\mathbf{x}_i \in \mathbb{R}^n|i=1,2,\ldots N\}$
in the embedding space.
The rows of the trajectory matrix correspond to events occurring at the
same time, while each column denotes an individual time series.
The oldest measurements on the system appear in the first row, whereas
the most recent ones appear in the last row.
In other words, applying the first lag to every series in $\mathbf{X}$,
then the second lag to every series in $\mathbf{X}$, and so forth and so
on, one obtains a sequence of lagged vectors to describe the whole
evolution trajectory of the measured system.

All these properties can be expressed in terms of the following
trajectory (or embedding) matrix
\begin{widetext}
\begin{equation}
\mathbf{X}
=
N^{-1/2}
\begin{bmatrix}
               x(t)	    & x(t-1)     & x(t-2)     & \dots  & x(t-n)     \\
               x(t-1)   & x(t-2)     & x(t-3)     & \dots  & x(t-1-n)   \\
               x(t-2)   & x(t-3)     & x(t-4)     & \ldots & x(t-2-n)   \\
               \vdots   & \vdots     & \vdots     & \ddots & \vdots     \\
               x(t-N_s)	& x(t-N_s-1) & x(t-N_s-2) & \dots  & x(t-N_s-n)
\end{bmatrix} \, ,
\label{eq:traj-matrix}
\end{equation}
\end{widetext}
where $N^{-1/2}$ is a convenient normalization factor and $n$ is the
embedding dimension.
This $(N-n)\times (n+1)$ matrix contains the complete history of the
measured dynamical system in the space of all $n$-element patterns in
the embedding space $\mathbb{R}^n$ and can be considered as a linear
map from $\mathbb{R}^n$ to $\mathbb{R}^N$ (see \cite{Broomhead:1985}
for further discussion).

\subsection{Attractor reconstruction}
\label{subsec:TFF-attractor}
Let us now specify the above considerations by displaying the time
series used in our TFF analysis.
We construct the basic series for the CLEO measurement
\cite{Gronberg:1997fj} from the data given in
Table \ref{tab:ff-values-table} by selecting events
at $t=Q^2$ values in steps of approximately 1~GeV$^2$.
This gives rise to the first column of the embedding matrix for
CLEO (label C), viz.,
\begin{equation}
  \mathcal{F}(t)^\text{C}
=
 [
   0.121, 0.151, 0.132, 0.154,
   0.155, 0.148, 0.167
 ]^T \, .
\label{eq:CLEO-basic}
\end{equation}
Shifting this series by the first lag, we get the second column, i.e.,
the lag-one-series
\begin{equation}
  \mathcal{F}(t-1)^\text{C}
=
 [
   0.151, 0.132, 0.154,
   0.155, 0.148, 0.167
 ]^T
\label{eq:CLEO-lag-one}
\end{equation}
and by shifting it by the second lag, we obtain the third column
(the lag-two series)
\begin{equation}
  \mathcal{F}(t-2)^\text{C}
=
 [
   0.132, 0.154,
   0.155, 0.148, 0.167
 ]^T \, .
\label{eq:CLEO-lag-two}
\end{equation}
\begin{figure}[h]
\includegraphics[width=0.40\textwidth]{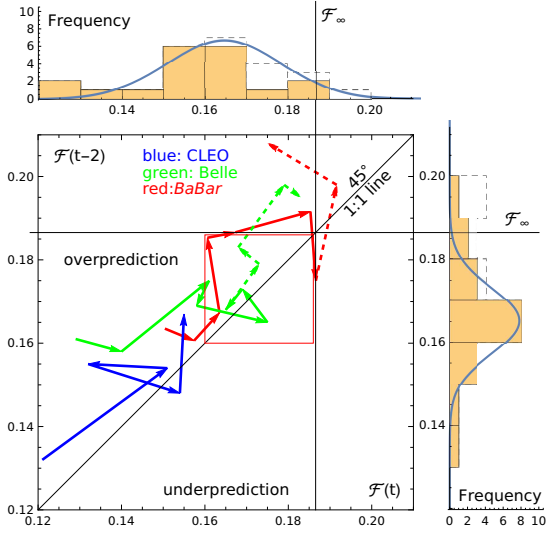}
\caption{
\label{fig:xz-scatter-histo}
2D projection of the reconstructed attractor in terms of
lagged coordinate vectors with respect to
$\mathcal{F}(t)$ and
$\mathcal{F}(t-2)$.
The same designations as in Fig.\ \ref{fig:xy-scatter-histo} are used.
}
\end{figure}
We then construct the attractor portrait in terms of 3D vectors on the
trajectory, denoted by the symbol $\bullet$,
from the first five rows of these three columns in the form
\begin{equation}
\bullet
=
 [\mathcal{F}(t), \mathcal{F}(t-1), \mathcal{F}(t-2)] \, ,
\label{eq:vector-form}
\end{equation}
while elements in the remaining rows are lost in the lagging process.

\begin{figure}[t]
\includegraphics[width=0.40\textwidth]{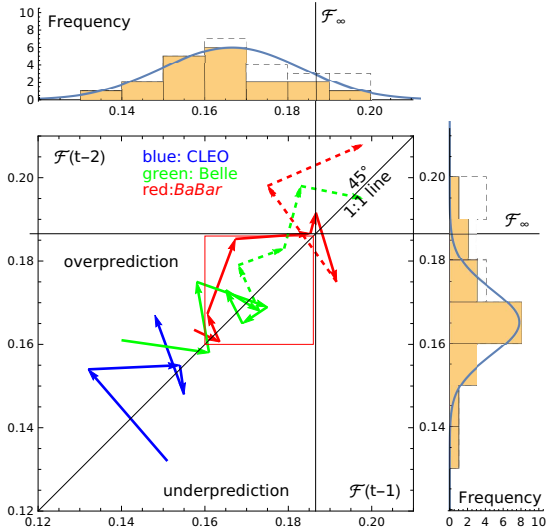}
\caption{
\label{fig:yz-scatter-histo}
2D projection of the reconstructed attractor in terms of
lagged coordinate vectors with respect to
$\mathcal{F}(t-1)$ and
$\mathcal{F}(t-2)$.
The same designations as in Fig.\ \ref{fig:xy-scatter-histo} are used.
}
\end{figure}

\begin{figure}[th]
\includegraphics[width=0.4\textwidth]{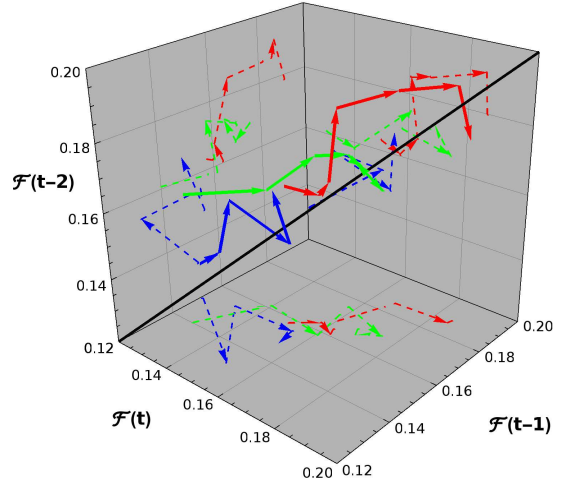}
\caption{
Attractor reconstruction in a 3D embedding phase space
using the same time series as in the previous 2D plots
related to three different sets of data and using the
designations explained in Fig.\ \ref{fig:xy-scatter-histo}.
The dashed lines show the footprints of the two-dimensional
projections displayed in
Figs.\ \ref{fig:xy-scatter-histo}, \ref{fig:xz-scatter-histo},
and \ref{fig:yz-scatter-histo}, while the additional vectors
shown there by broken lines are not included.
The black solid line marks the $1:1$ line at $45^\circ$.
\label{fig:Packard-Takens_data-3D}
}
\end{figure}

The basic series from the \textit{BABAR} (label B) \cite{Aubert:2009mc}
and the Belle (label b) \cite{Uehara:2012ag} data,
displayed in Table \ref{tab:ff-values-table},
are assembled analogously to obtain
\begin{eqnarray}
  \mathcal{F}(t)^\text{B}
&& \!\!\!\!\!\! =
 [
   0.150, 0.157, 0.164, 0.161, 0.167, 0.185,
\\ \nonumber
&&  ~ 0.187,
\underline{0.192}, \underline{0.175}, \underline{0.198}, \underline{0.208}
 ]^T \, ,
\label{eq:BABAR-basic}
\end{eqnarray}
\begin{eqnarray}
  \mathcal{F}(t)^\text{b}
&& \!\!\!\!\!\! =
 [
   0.129, 0.140, 0.161, 0.158, 0.175, 0.169,
\\ \nonumber
&&  ~ 0.165,
  \underline{0.173}, \underline{0.168}, \underline{0.179},
  \underline{0.183}, \underline{0.198}, \underline{0.195}
 ]^T \, ,
\label{eq:Belle-basic}
\end{eqnarray}
respectively.
From these two basic series, we deduce the corresponding lag-one
and lag-two series and then create from the \textit{BABAR} data
seven regular and two approximate 3D vectors, whereas from the
Belle data we obtain seven regular 3D vectors and four approximate
ones.
The approximate vectors in each case have a sampling time
$1<\tau_s<2.0$~GeV$^2$ and refer to the underlined numbers in
the corresponding basic series.
All data of both experiments above 15.95~GeV$^2$ (\textit{BABAR})
and 22.24~GeV$^2$ (Belle) cannot be included in the corresponding
basic series because they were measured at distant momentum values
much larger than the sampling time $\tau_s=1$~GeV$^2$.
Thus, they are excluded by the embedding procedure and not arbitrarily.
Unfortunately, even the approximate vectors correspond to momenta
below 20~GeV$^2$, notably,
$Q^2=12.71$~GeV$^2$ (ninth \textit{BABAR} vector) and
$Q^2=16.96$~GeV$^2$ (eleventh Belle vector).
Therefore, the measurements at momenta larger than these values, where
one would expect a scaling behavior of the TFF starting to emerge,
could not be taken into account in the considered attractor
reconstruction.
However, this is not a deficiency of the method but the consequence
of sparse measurements in this $Q^2$ region.

The two-dimensional (2D) projections of the phase-space
reconstruction are shown in Figs.\ \ref{fig:xy-scatter-histo},
\ref{fig:xz-scatter-histo}, \ref{fig:yz-scatter-histo} (with
designations explained in the first of them).
Each trajectory represents a separate phase portrait of the attractor
related to a particular measurement of $\mathcal{F}(Q^2)$ according
to Eq.\ (\ref{eq:trajectory}).
A dashed lining is used for the approximate vectors to indicate that
there are some missing jags along this trajectory owing to the
fact that the sampling time is larger than $\tau_s=1$~GeV$^2$.
In other words, the event collection in this part of the basic series
entails a certain imprecision in the structure of the trajectory.

The scatter-plot emerging from the CLEO data is shown in blue color and
starts close to the bottom and ends at the center.
The \textit{BABAR} trajectory (red lines) occurs at the center and
extends to the far-end of the displayed time series,
while the Belle trajectory (in green color) crosses the other two
in between as it climbs.
No experimental errors are included in these figures.
For illustration, we combine these 2D projections of the considered
time series into a 3D graphics shown in
Fig.\ \ref{fig:Packard-Takens_data-3D}.
Note that the approximate vectors are not included here.
The broken lines denote instead the various 2D projections
corresponding to Figs.\ \ref{fig:xy-scatter-histo},
\ref{fig:xz-scatter-histo}, \ref{fig:yz-scatter-histo}.

From these figures we observe that the structure of the data
attractors has been sufficiently resolved: not too smooth (ordered),
not too jagged (disordered)---just right to recognize a
deterministically generated pattern.
This provides justification for the choice of the employed embedding
parameters.
Moreover, though the reconstructed attractors, emerging from each of
these times series, show some idiosyncratic structure, they are
composed of vectors clustering rather close to each other with a
frequency peak in the range
\begin{equation}
  \mathcal{F}(Q^2)\in [0.16-0.17]~\text{GeV} \, ,
\label{eq:attractor}
\end{equation}
occurring in the momentum region
\begin{equation}
  Q^2\in [9-11]~\text{GeV}^2 \, ,
\label{eq:attr-mom-range}
\end{equation}
as quantified by the histograms.
As one can see from Table \ref{tab:ff-values-table}, TFF values around
this estimate have been measured by all three considered experiments
in the momentum range $Q^2\in [6-10]$~GeV$^2$.
Counting all 12 measurements in the range $[6.47-10.48]$~GeV$^2$, we
get a statistical average of $\mathcal{F}\approx 0.167$~GeV in good
agreement with the attractor value in the histograms.
Including into the reconstruction procedure the displayed approximate
vectors, the histograms are shifted closer to the asymptotic limit
$\mathcal{F}_\infty$ so that the distributions become negatively skewed
towards this value, while the estimated TFF value slightly increases.
Ultimately, the geometry of the trajectory segments inside the
asymptotic attractor area becomes irrelevant, because the system has
reached an equilibrium state of its dynamics close to the fixed point
$\mathcal{F}_\infty$
with the state vectors pointing in opposite directions at almost equal
rates.

Based on this reasoning, we argue that the obtained attractor
structure,
cf. Eqs.\ (\ref{eq:attractor}), (\ref{eq:attr-mom-range}),
where all particular trajectories have repetitive vector occurrences
to neighboring states,
represents a prodromal portrait of the true asymptotic attractor that
would emerge if we would have at our disposal a finer partition of
measurements in steps of 1~GeV$^2$ in the range between 10~GeV$^2$ and
25~GeV$^2$ to improve the convergence of the bootstrapping procedure
to the correct limit. 
We estimate that to get a faithful attractor reconstruction, we need
from a future experiment a series of TFF measurements starting, say, at
4.48~GeV$^2$ and continuing up to 25.48~GeV$^2$ in steps of 1~GeV$^2$.
This way, we would obtain in total 20 3D state vectors for the
attractor reconstruction.
Assuming experimental errors similar to or smaller than those of the
Belle measurements, this attractor portrait would suffice to establish
the asymptotic limit of the TFF already below 25~GeV$^2$.
That is to say, the attractor not only provides a shortcut to abbreviate
the experimental efforts, it can also be used as a diagnostic tool to
sort out events that contradict scaling \emph{above} this momentum.
On the other hand, if the described scenario will not be confirmed,
the odds are stacked against the experimental observation of scaling in
the TFF.
\begin{figure}[th]
\includegraphics[width=0.392\textwidth]{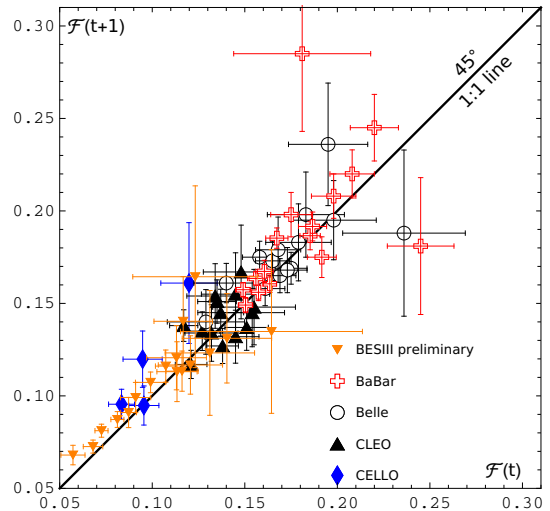}
\caption{2D attractor reconstruction
from all existing data sets using each of them as a basic time
series with varying sampling-time intervals.
The state space reconstruction fails because the attractor remains
hidden in the ``width'' of the data distribution along the $1:1$ line
at $45^\circ$.
Those data, which are compatible with an asymptotic scaling of the TFF,
show a strong positive correlation, whereas the auxetic
\textit{BABAR} data above $\sim 10$~GeV$^2$ and the
Belle outlier at 27.33~GeV$^2$ (see Table \ref{tab:ff-values-table})
are distributed according to a significant \textit{negative}
correlation.
\label{fig:Packard-Takens_data-2D}
}
\end{figure}

We postpone further discussion of these figures to the next
section and consider the possibility of applying the method of time
delays to all data (CELLO, CLEO, \textit{BABAR}, Belle,
together with their experimental errors
(see Table \ref{tab:ff-values-table})
using for each of them an unevenly sampled basic series from the
recorded measurements as they appear in Table
\ref{tab:ff-values-table}.
We also include the BESIII preliminary data
\cite{Redmer:2018uew} as they appear, i.e., with a varying sampling
time $\tau_s$.
We only consider the correlation of this ensemble of basic
time series with the common ``lagged-one'' time series
$\mathcal{F}(t-1)$, i.e., we fix the embedding dimension to
$n=2$.
The results are plotted in Fig.\ \ref{fig:Packard-Takens_data-2D}.
Despite the application of an inaccurate phase-space reconstruction,
resulting from the extremely short sampling times and varying
lags,
there are some striking observations from this figure.
First, as expected, the attractor structure fails to unfold, with all
state vectors being ``buried'' inside the (noisy) data width of
the $1:1$ line showing in toto a positive correlation.
This behavior agrees with the global trend of the attractors
determined before using a strict embedding procedure.
Second, the linear regression trend is especially accentuated in
the case of the preliminary BESIII data at $Q^2$
values below 1.5~GeV$^2$, which are very close to each other and
thus induce a strong autocorrelation.
On the other hand, the large total error of these data at the upper
end of $Q^2$ up to 3.1~GeV$^2$ does not allow to draw reliable conclusions.
Third, in contrast to this overall trend of the entirety of the
analyzed data, there is a segregated group of states forming a pattern,
which, as a whole, exhibits a \emph{negative} correlation.
Interestingly, this pattern pertains exactly to those
\textit{BABAR} data points that deviate from the scaling limit of
pQCD (see Fig.\ \ref{fig:linear-scaled-TFF}).
The Belle outlier at $Q^2=27.33$~GeV$^2$
(see Table \ref{tab:ff-values-table}) also belongs to this group.
These opposing tendencies of the system trajectories cannot be
attributed to a common dynamical mechanism for the evolution of the
measured system in its state space, pointing to an intrinsic
incompatibility within the data.
Remarkably, this inconsistency cannot be inferred from
Fig.\ \ref{fig:linear-scaled-TFF} in the statistical sense
because, as shown in \cite{Uehara:2012ag,Stefanis:2012yw}, the relative
deviation between the \textit{BABAR} and Belle data fits does not
exceed $1.5\sigma-2\sigma$.

\section{LCSR predictions versus phase-space reconstruction}
\label{sec:pred-forecasts}
In this section we compare the TFF predictions obtained
within QCD in
Sec.\ \ref{sec:analysis-theory} with the attractor phase portrait
extracted from the data in the previous section.
The strategy is to identify those particular features of the calculated
TFF that provide agreement or disagreement with the determined
attractor.
To a great extent, this evaluation integrates and expands our previous
analysis of the typical pion DAs considered in
Table \ref{tab:pion-DAs-a6}
and Figs.\ \ref{fig:linear-scaled-TFF}, \ref{fig:omega-scaled-TFF}.

To this end, let us first provide a brief quantitative assessment of
the TFF calculation within our QCD-based LCSR approach.
Using collinear factorization, the leading-twist part of the TFF at
the NNLO of the perturbative expansion reads
(see Sec.\ \ref{sec:analysis-theory} for the explicit expressions)
\begin{widetext}
\begin{eqnarray}
  F^{\gamma^*\gamma^*\pi^0}(Q^2,q^2)
\! &\! = \! & \!
  N_\text{T} \Bigg[
              \underbrace{T_\text{LO}}_{(+)} + a_s(\mu^2)\underbrace{T_\text{NLO}}_{(-)}
            + a^2_s(\mu^2)\bigg(
                                  \underbrace{T_{{\rm NNLO}_{\beta_0}}}_{(-)}
                                + \underbrace{T_{{\rm NNLO}_{\Delta V}}}_{(-)}
                                + \underbrace{T_{{\rm NNLO}_{L}}}_{(0)}
                                + \underbrace{T_{{\rm NNLO}_{c}}}_{(?)}
                          \bigg)
                         + \ldots
      \Bigg] \otimes \varphi_\pi^{(2)}(x,\mu^2)
\nonumber \\
~~~~~~~~~~& + \! & \! \mathcal{O}\left(\frac{\delta^2}{Q^4}\right)
\label{eq:full-TFF}
\end{eqnarray}
with indications showing the sign of these contributions.
The label $(?)$ marks the only still uncalculated term.
It is the source of the major theoretical uncertainties illustrated
in Figs.\ \ref{fig:linear-scaled-TFF} and \ref{fig:omega-scaled-TFF}
in terms of the wider blue shaded band enveloping the green one.
The other NNLO contributions are included.
Employing the LCSR formalism, we obtain the TFF for one highly virtual
and one quasireal photon in the form
\begin{equation}
  Q^2F^{\gamma*\gamma\pi^0}(Q^2)
=
  F^\text{tw-2}(Q^2) + F^\text{tw-4}(Q^2) + F^\text{tw-6}(Q^2) \, ,
\label{eq:TFF-twists}
\end{equation}
\end{widetext}
where
\begin{equation}
  F^\text{tw-2}(Q^2)
=
  F_{0}(Q^2) + \sum_{n} a_n(Q^2)F_n(Q^2) \, .
\label{eq:TFF-tw-2}
\end{equation}

Referring to the above equations, we now summarize the key ingredients
of the LCSR analysis taking also into account relevant results from
previous investigations.
\begin{itemize}
\item
The Tw-4 term is negative, whereas the Tw-6 contribution is positive.
Both are included explicitly as explained in
Sec.\ \ref{sec:analysis-theory}.
\item
NLO evolution with heavy-quark thresholds provides suppression
that depends on the heavy-quark masses and the amount of the
Gegenbauer coefficients included in the conformal expansion of the
pion DA, see Sec.\ \ref{sec:analysis-theory} and
App.\ \ref{sec:global-NLO-evolution}.
\item
Consideration of a finite virtuality of the quasireal photon
also leads to suppression \cite{Stefanis:2012yw}.
This effect is not universal; it depends on the experimental set-up
and is of minor importance for our present study.
Therefore, it is not included.
\item
As a rule, DAs with suppressed tails $x=0,1$ tend to decrease the
size of the TFF, while those with endpoint enhancement tend to
increase it.
For a quantitative treatment of these issues, we refer to
\cite{Mikhailov:2010ud,Stefanis:1998dg,Stefanis:2014yha,Stefanis:2015qha}.
In particular, the interplay between the ``peakedness'' of a DA at
$x=0.5$ and the ``flatness'' or enhancement of its tails at $x=0,1$
can be quantified in terms of the kurtosis statistic
$
 \beta_2[\varphi]
=
 \langle \xi^4\rangle/\left(\langle \xi^2\rangle\right)^2
$, see \cite{Stefanis:2015qha}.
\item
The signs and magnitudes of the Gegenbauer coefficients also have a
strong effect on the overall size of the TFF.
This becomes evident by recalling Eq.\ (\ref{eq:inv-mom}) for the
inverse moment.
It implies that the Gegenbauer coefficients have to
balance each other in such a way as to provide just the right amount
of enhancement relative to the asymptotic DA.
For example, the coefficients $a_2$ and $a_4$ of the BMS DA
and the platykurtic DA have comparable magnitudes but opposite signs
(see Table \ref{tab:pion-DAs-a6}), whereas higher-order coefficients
are marginal and contribute mainly to the theoretical uncertainties,
so that the negative contributions mentioned in the previous items
amount to a reduction of the total value of the TFF just to the
gross size of the data, except the auxetic ones.
There is a variation in the rate and extent of the influence of
the Gegenbauer coefficients as one can see for some other DAs
from Table \ref{tab:pion-DAs-a6} in comparison
with Fig.\ \ref{fig:linear-scaled-TFF}.
\item
Among the considered pion DAs, the platykurtic model
\cite{Stefanis:2014nla} has the following advantages:
(i) It has only two conformal coefficients $a_2$ and $a_4$,
with all higher coefficients being compatible with zero.
This enables ERBL evolution at the two-loop level including heavy-quark
thresholds (App.\ \ref{sec:global-NLO-evolution}).
(ii) It amalgamates by construction endpoint suppression
(via $\lambda_{q}^{2}=0.45$~GeV$^2$)
with unimodality (like in DCSB DAs).
(iii) It gives $\varphi_{\pi}^{(2)\text{pk}}(x=0.5,\mu_2)=1.33$
and thus satisfies the constraint $1.2\pm 0.3$ calculated with LCSRs
in \cite{Braun:1988qv}.
(iv) It yields
$\langle x^{-1}\rangle(\mu_2)=3.13$ (Table \ref{tab:pion-DAs-a6}),
which is sufficient to accurately describe all data from low to high
$Q^2$, provided the latter are compatible with scaling.
(v) It resides inside the asymptotic regime of the determined
attractor
cf. Eqs.\ (\ref{eq:attractor}), (\ref{eq:attr-mom-range}),
once it reaches the value $\mathcal{F}(Q^2)=0.16$
around $Q^2 \lesssim 11$~GeV$^2$.
(vi) It complies within the margin of error with the new lattice
results at NNLO and NLO \cite{Bali:2019dqc} for
$a_2$ at $\mu_2$,
keeping in mind that the NLO evolution procedure
with heavy-quark thresholds induces a stronger reduction of the
initial value (Table \ref{tab:pion-DAs-a6}).
All these features ensue from the applied construction procedure
\cite{Stefanis:2015qha} and are not the product of a fit to any data.
\item
Best agreement with the \textit{BABAR} auxetic data can be achieved by
using flat-type DAs \cite{Radyushkin:2009zg,Polyakov:2009je}, though
also DAs with an inverse hierarchy of several Gegenbauer coefficients,
like model II in \cite{Agaev:2010aq}, may also provide conforming
predictions.
Flat-type DAs overestimate both, the CLEO and the Belle data
(see \cite{Bakulev:2012nh,Agaev:2012tm}).
\end{itemize}

\begin{figure}[th]
\includegraphics[width=0.48\textwidth]{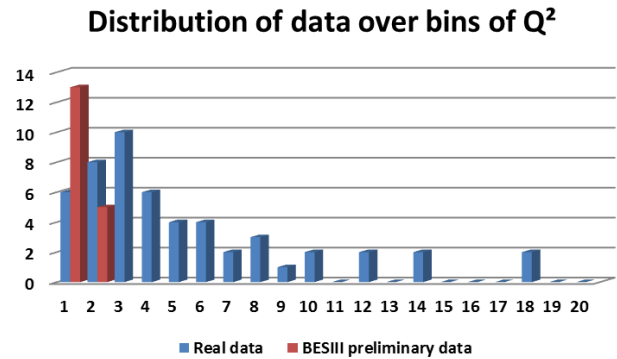}
\caption{Histograms of assembled data from
Table \ref{tab:ff-values-table} in bins of $Q^2$ in the range between
0 and 40 GeV$^2$ (20 bins in total).
The momentum increment between markers on the $x$ axis is 2~GeV$^2$.
The $y$ axis shows the frequency, i.e., the total number of data points
(blue color) in each bin by combining all available measurements
(CELLO \cite{Behrend:1990sr},
CLEO \cite{Gronberg:1997fj},
\textit{BABAR} \cite{Uehara:2012ag}, and
Belle \cite{Aubert:2009mc}).
The preliminary data of BESIII \cite{Redmer:2018uew}
below bin 3 are shown separately in red color.
Only $35\%$ of the data occurs above bin 5, i.e., above 10~GeV$^2$.
\label{fig:data-distribution}
}
\end{figure}

We now turn our attention to the topology-based data analysis.

First, using for each of the CLEO, \textit{BABAR}, and Belle data sets
a basic time series with a sampling interval $\tau_s=1$~GeV$^2$, we
reconstructed the state space of the TFF in terms of lagged coordinate
vectors.
More importantly,
in Figs.\ \ref{fig:xy-scatter-histo}, \ref{fig:xz-scatter-histo},
\ref{fig:yz-scatter-histo},
we identified and quantified in terms of histograms a common restrained
region (red square) within this space, where the \textit{BABAR} and
Belle trajectories show both an aggregation of state vectors around the
value
$\mathcal{F}(Q^2)=0.165\pm 0.005$~GeV
that appears in the momentum range
$Q^2\in [9-11]$~GeV$^2$.
We argued that this attractor regime is a transient version of the true
long-term attractor that would emerge closer to $\mathcal{F}_\infty$,
if a more dense set of data between 10 and 25~GeV$^2$ would have been
included in the analysis.
This $Q^2$ range was inefficiently covered in all experiments so far.

This becomes evident from Fig.\ \ref{fig:data-distribution}, where we
show the experimental data, given in Table \ref{tab:ff-values-table},
plotted as a histogram across the values of $Q^2$ from 0 to 40~GeV$^2$.
The bins show combined data of different experiments within successive
intervals of 2~GeV$^2$ each numbered from 1 to 20, starting from the
interval
$1:[0-2]$~GeV$^2$ up to the interval $20:[38-40]$~GeV$^2$.
There are in total 52 points composed as follows:
CELLO (5), CLEO (15), \textit{BABAR} (17), and Belle (15).
For the sake of completion, the 18 preliminary data
points of BESIII \cite{Redmer:2018uew} have also been included using a
different color to distinguish them from final data
published in peer-reviewed journals.
As one observes from this figure, the data do not form a uniform
distribution over $Q^2$.
In fact, over 50 percent of the events occur below
bin~$5:[10-12]$~GeV$^2$.
The $Q^2$ intervals above 20~GeV$^2$ are only scarcely populated
and some of the data in this high-end $Q^2$ regime bear rather large
errors (see Table \ref{tab:ff-values-table} and
Fig.\ \ref{fig:linear-scaled-TFF}).

Second, from the approximate embedding procedure shown in
Fig.\ \ref{fig:Packard-Takens_data-2D},
we observed a conflicting behavior of the data concerning the states
of the TFF in its phase space.
While most data points, encompassing the preliminary BESIII,
CELLO, CLEO, Belle, and \textit{BABAR} below 10~GeV$^2$, describe TFF
states that follow in round terms a positive correlation pattern along
the diagonal, the high-end \textit{BABAR} data together with the Belle
event at 27.33~GeV$^2$ are found to be negatively correlated.
Data from new experiments may contribute to the clarification of this
observed discrepancy.
In this context, we mention that this observation is in line with
the statistical analysis carried out in \cite{Stefanis:2012yw}
in which we investigated the possibility to predict the trend of
the Belle and the \textit{BABAR} data from one another.
We found that both popular parametrizations, a dipole and a power-law
function can fit the Belle data in a satisfactory way, while only
the power fit works well for the \textit{BABAR} data.
However, using the corresponding parameters of these fit functions,
determined for each of these sets, one cannot reproduce the other with
an acceptable statistical precision.
The inclusion of the CELLO and CLEO data into this fitting procedure
does not modify this finding.
The conclusion drawn in \cite{Stefanis:2012yw} was that the Belle and
\textit{BABAR} data segregate into two separate classes that cannot
be fitted simultaneously.

Third, from Table \ref{tab:ff-values-table}, we observe that the TFF
value calculated with the BMS DA
(or equivalently the platykurtic one)
does not change appreciably after $Q^2=10.48$~GeV$^2$, where it reaches
the value 0.161~GeV.
This implies that the corresponding phase-space trajectory will enter
at this scale the red square determined in Figs.\
\ref{fig:xy-scatter-histo},
\ref{fig:xz-scatter-histo},
\ref{fig:yz-scatter-histo}
and then stay within it, merging practically with the diagonal after
some point.
Thus, this TFF prediction agrees with the existence of the
reconstructed attractor and supports a saturating behavior of the TFF
starting around [9-11]~GeV$^2$.
On the other hand, TFF predictions that exceed the asymptotic limit
within the range of the available data, e.g., the result obtained with
the DSE-DB DA, will give rise to phase-space trajectories following the
direction of the diagonal after exiting the attractor portrait
(the red square) and continuing to grow slowly at some variable
distance from it.
In fact, the histograms in that case are approximately uniform with no
distinctive maximum at some value.
Finally, TFF predictions that are tailored to reproduce the auxetic
\textit{BABAR} data in terms of a flat pion DA, will cross the
attractor residence within the red square and then diverge from it
towards increasingly larger values along jagged trajectories.
This just reflects the feature that the fluctuations away from the
diagonal can be much larger than the average direction of the evolution
trajectory on the attractor.
Consequently, such predictions are not compatible with the existence of
an attractor within the TFF data showing up in the range
$[9-11]$~GeV$^2$ and, therefore, they contradict the onset of
QCD scaling in this $Q^2$ domain.

\section{Conclusions}
\label{sec:concl}
This section summarizes the benchmarks of our analysis and presents our
conclusions.

In this work we scrutinized the single-tagged process
$e^+e^- \rightarrow e^+e^-\pi^0$
in terms of the pion-photon transition form factor
$\gamma^*\gamma\rightarrow\pi^0$,
described by the quantity
$F^{\gamma^*\gamma\pi^0}(Q^2)$.
Observations of this exclusive process can provide insight into the
dynamical characteristics of the electromagnetic $\pi^0$ vertex and its
microscopic explanation at the quark-gluon level using particular
theoretical formalisms or models.
To obtain reliable predictions and include their intrinsic
uncertainties, we employed a LCSR-based scheme which is embedded into
the collinear factorization framework of QCD and the twist expansion.
This dispersive approach allows the systematic inclusion into the
spectral density of perturbative radiative corrections together with the
contributions from higher twists.
Moreover, the obtained expression for the TFF retains its validity also
in the case of a quasireal photon emitted from the untagged electron.
The approach can be used with various pion DAs in terms of their
conformal expansions, thus facilitating the inclusion of ERBL
evolution.

In the presented analysis, we included perturbative contributions to the
hard-scattering amplitude up to the NNLO, except a single term which is
still unknown, see (\ref{eq:T}).
The used spectral density also includes the twist-four and
twist-six corrections.
In our predictions the binding effects of the pion were taken into
account in the form of various pion DAs.
We used two twist-two DAs derived with QCD sum
rules employing nonlocal condensates \cite{Mikhailov:1986be}.
One family of DAs has a bimodal profile and suppressed tails at
$x=0,1$ \cite{Bakulev:2001pa} ($\lambda_{q}^2=0.4$~GeV$^2$).
Allowing for a slightly larger (but still admissible) average vacuum
quark virtuality $\lambda_{q}^2=0.45$~GeV$^2$, one can obtain a DA
with a short-tailed platykurtic profile \cite{Stefanis:2014nla}.
The TFF computed with these DAs at the momentum values probed
experimentally are given in Table \ref{tab:ff-values-table} together
with the chief theoretical uncertainties discussed in
Sec.~\ref{sec:analysis-theory}.

To obtain a variety of TFF predictions, we also employed pion DAs
obtained in other approaches, for instance, the DAs with enhanced
endpoint regions from DSE-based calculations
\cite{Chang:2013pq,Raya:2015gva} or
AdS/QCD \cite{Brodsky:2011xx}.
Predictions from some other models have also been included, see
Fig.\ \ref{fig:linear-scaled-TFF}.
To connect the calculated TFF predictions at the initial scale (either
$\mu_1=1$~GeV or $\mu_2=2$~GeV) to the measurements at higher momentum
scales, we employed NLO evolution that contains an arbitrary number of
conformal coefficients and takes into account heavy-quark thresholds
(see App.\ \ref{sec:global-NLO-evolution}).
The statistical measures to quantify the interplay of the
``peakedness''
(in terms of the second moment or $a_2$)
and the ``tailedness''
(by means of the fourth moment or $a_4$) of these DAs were
discussed in connection with Table \ref{tab:pion-DAs-a6}.

To analyze the asymptotics of the TFF predictions, we invented and
used a new quantity which measures the deviation of the TFF value
from the asymptotic limit set by pQCD, see
Eq.\ (\ref{eq:scaling-rate})
and Fig.\ \ref{fig:omega-scaled-TFF}.
The upshot of these considerations can be encapsulated in the
following statement.
Pion DAs with enhanced tails tend to increase the magnitude of the TFF
to a level above the asymptotic limit
$\mathcal{F}_\infty=\sqrt{2}f_\pi$,
while the considered DAs with suppressed endpoint regions $x=0,1$
yield predictions that approach this limit from below without
crossing it.
Both types of predictions are, therefore, compatible with the scaling
property of pQCD at asymptotic momentum scales and therefore disagree
with the auxetic \textit{BABAR} data points above 10~GeV$^2$.
We didn't discuss model calculations attempting to explain this data
behavior and refer to our previous dedicated analysis in
\cite{Bakulev:2012nh} and references cited therein.

The QCD-based calculations were supplemented by a
topology-based data
analysis, carried out in Sec.\ \ref{sec:data-analysis}.
We showed that topological embedding subject to Taken's theorem
provides a mathematical scaffolding to carry out a nonlinear
time-series analysis of the observed TFF data to unveil the underlying
dynamics without appealing to any theoretical formalism or model.
The key components of the method to reconstruct the state space of
the system (the TFF) were described and the embedding matrix for the
delayed time series vectors was worked out.
The intrinsic and practical limitations of the method were also pointed
out.
Using appropriate embedding parameters, we determined the phase
portrait of two attractors in 3D space, one related to the data of the
Belle experiment and the other pertaining to \textit{BABAR}.
The controversial branch of the auxetic \textit{BABAR} data above
10~GeV$^2$ was excluded by the applied embedding parameters---not
arbitrarily.
Nevertheless, both attractor structures enter a common area of phase
space, where they have a maximum of state vector occurrences
around the value $\mathcal{F}(Q^2)=0.165\pm0.005$~GeV
emerging at scales in the range $Q^2\in [9-11]$~GeV$^2$
and marking the onset of asymptotic scaling.
This dynamical characteristic was quantified in terms of histograms
in Figs.\ \ref{fig:xy-scatter-histo}, \ref{fig:xz-scatter-histo},
\ref{fig:yz-scatter-histo}.
We argued that this generic attractor portrait provides a shadowing reflection
of the true asymptotic attractor to be determined from future
experiments, e.g., Belle-II.
We claimed that the final arrangement of the data-driven state vectors
can be revealed by the outcome of measurements from
$Q^2=10.48$~GeV$^2$ to $Q^2=25.48$~GeV$^2$
(like in the Belle experiment)
with a fixed increment of 1~GeV$^2$.
We encourage the Belle-II Collaboration to design the data acquisition
of their experiment accordingly.

Understanding the phase-space structure of the data on the pion-photon
transition would be an essential step towards determining the
asymptotic behavior of the TFF that follows from the basic principles
of QCD.
We have shown that endpoint-suppressed pion DAs yield predictions which
enter the attractor regime and then remain inside it.
By contrast, TFF predictions derived with endpoint-enhanced DAs
cross the asymptotic attractor area and then leave it again towards
larger TFF values.
Thus, a validated and accurate phase-state portrait of the attractor
will provide a reliable diagnostic tool to select the most appropriate
type of the pion DA.
Combining it with the new more reliable lattice results of the
RQCD Collaboration \cite{Bali:2019dqc} for the second conformal
coefficient $a_2$, can reduce the variation of the DA profile even
further (see the previous section).
To this end, the determination of $a_4$ would be extremely useful but
difficult to realize on the lattice \cite{Bali:2019dqc}.
In this sense, the attractor represents from the experimental side, a
shortcut because it provides the possibility to establish the
asymptotic structure of the TFF attractor and the observation of QCD
scaling at much lower momentum values than anticipated until now, thus
avoiding overemphasis of solitary data points with unconfirmed accuracy
at much higher $Q^2$.
Implicit in this statement is the optimistic perspective that the
asymptotic regime of the TFF can be reached in a momentum range
accessible to experiments.
The outlined methodology provides the conceptual tools to obtain tangible
results in this direction.

\acknowledgments
I would like to thank Sergey Mikhailov and Alex Pimikov for collaboration
on many issues addressed in this work and for insightful comments on the
manuscript.
Special thanks are due to Alex Pimikov for handling the numerical
computations and their graphical representation.
I am grateful to Gunnar Bali for useful remarks.

\bigskip
\begin{appendix}
\section{NLO evolution of the pion DA with an arbitrary number of
Gegenbauer harmonics and including heavy-quark thresholds}
\label{sec:global-NLO-evolution}
In this Appendix
(done in collaboration with S.\ V.\ Mikhailov and A.\ V.\ Pimikov)
we discuss the NLO (i.e., two-loop) ERBL
evolution of the pion DA with an arbitrary number of Gegenbauer
coefficients and taking into account heavy-quark
flavors (also known as global QCD scheme, see, e.g.,
\cite{Bakulev:2012sm}).
This scheme employs the global coupling
$\alpha_{s}^\text{glob}(Q^2,\Lambda_{N_f}^2)$
that depends on the number of flavors $N_f$ through the QCD
scale parameter $\Lambda_{N_f}$.

This procedure was used in this work to derive the results given
in Tables \ref{tab:pion-DAs-a6} and \ref{tab:ff-values-table}
and obtain the predictions shown in Figs.\ \ref{fig:linear-scaled-TFF}
and \ref{fig:omega-scaled-TFF}.
It takes into account the heavy-quark thresholds and thus requires
the matching of the strong coupling in the Euclidean region of $Q^2$ at
the corresponding heavy-quark masses when one goes from
$N_f \to N_{f}+1$.
Note that the dependence on $N_f$ in Appendix D of
\cite{Bakulev:2002uc},
which provided the basis for the NLO evolution of the pion DA in our
earlier works, was ignored assuming a fixed number of flavors.
The new scheme has already been used in our more recent investigations
\cite{Bakulev:2012nh,Stefanis:2012yw,Stefanis:2014yha,Mikhailov:2016klg},
but without exposing the underlying formalism.
This task will be accomplished here.
The NLO evolution of the pion DAs with two conformal coefficients
$a_2$ and $a_4$ at the initial scale $\mu^2\simeq 1$~GeV$^2$
and a varying number of heavy flavors has also been applied in
\cite{Bakulev:2004cu} (see Appendix D there).
Our technical exposition below extends this treatment to any number
of conformal coefficients and more heavy-flavor thresholds,
see \cite{Bakulev:2008td,Bakulev:2012sm} for
details and further references.

Let us start with a fixed number of flavors and supply some basic
formulas from \cite{Bakulev:2002uc}.
The ERBL evolution equation for the pion DA is given by
\begin{equation}
  \frac{d\,\varphi_{\pi}(x;\mu^2)}{d\,\ln\mu^2}
=
  V\left(x,u;a_s(\mu^2)\right)\underset{u}\otimes\varphi_{\pi}(u;\mu^2)
\label{eq:ERBL-eq}
\end{equation}
and is driven by the kernel
\begin{equation}
  V(x,y;a_s)
=
  a_s~ V_0(x,y)
  + a_s^2~ V_1(x,y)
  + \ldots
\label{eq:kernel}
\end{equation}
with $a_s=\alpha_s/(4\pi)$.

The eigenvalues $\gamma_{n}(a_s)$
and the one-loop eigenfunctions $\psi_{n}(u)$
are related to the kernel $V$ through
\begin{equation}
  \tilde{\psi}_{n}(x) \underset{x}\otimes V(x,u;a_s)\underset{u}\otimes \psi_{n}(u)
=
  - \gamma_n(a_s)\, ,
\label{eq:eigen}
\end{equation}
where
$
 \displaystyle \tilde{\psi}_{n}(x)
=
 2(2n+3)/\left[3(n+1)(n+2)\right] C^{3/2}_n(x-\bar{x})
$.
The explicit expressions for the anomalous dimensions $\gamma_{n}$
at one loop, $\gamma_0(n)$,
and two-loops, $\gamma_1(n)$,
in the expansion
$
 \gamma_{n}(a_s)
=
 \frac{1}{2}[a_s\gamma_{0}(n) + a_{s}^2\gamma_{1}(n) + \ldots]
$
can be found in Appendix D in \cite{Bakulev:2002uc}.

To perform the pion DA evolution, while ignoring quark-mass thresholds,
we make use of the evolution matrix $E$ with the components
$E_{nk}$.
Expanded over the basis $\{\psi_n\}$ of the Gegenbauer harmonics, this
matrix assumes the following triangular form \cite{Mikhailov:1985cm}
\begin{widetext}
\begin{eqnarray}
\label{eq:DA_EVO-glob}
  E_{nk}(N_f;Q^2,\mu^2)
&=&
  P(n,Q^2,\mu^2)\left[\delta_{nk}+ a_s(Q^2)
  \Theta(k-n>0)d_{nk}(Q^2,\mu^2)\right]\, , \\
d_{nk}(\mu^2,\mu^2) = 0 \, ,
\label{eq:e-norm}
\end{eqnarray}
where the coefficients
$d_{nk}(Q^2,\mu^2)$
will be defined shortly, and where $\mu^2$ and $Q^2$ refer to
the initial and observation scale, respectively.
The factor $P(n,Q^2,\mu^2)$ in Eq.\ (\ref{eq:DA_EVO-glob}) denotes the
diagonal part of the evolution matrix that dominates the
renormalization-group (RG) controlled evolution of the
$\psi_n$-harmonics in the conformal expansion
\begin{eqnarray}
  \varphi_\pi^\text{RG}(x,Q^2)
&=& \sum\limits_{n} a_{n}(\mu^2)\left\{ P(n,Q^2,\mu^2)
  \left[\psi_{n}(x) + a_{s}(Q^2)
  \sum\limits_{k>n} d_{nk}(Q^2,\mu^2) \psi_{k}(x)
  \right]\right\}\, .
\label{eq:DA_EVO}
\end{eqnarray}
Then, the diagonal part of the evolution exponential at the
two-loop level can be given explicitly,
\begin{eqnarray}
  P(n,Q^2,\mu^2)
&=&
  \exp
      \left[
            \int\limits_{a_{s}(\mu^2)}^{a_{s}(Q^2)}
            \frac{\gamma_n(a)}{\beta(a)}da
      \right]
\stackrel{2-\text{loops}}{\longrightarrow}
      \left[
            \frac{a_s(Q^2)}{a_s(\mu^2)}
      \right]^{\frac{\gamma_0(n)}{2b_0}}
      \left[\frac{1+c_1a_s(Q^2)}{1+c_1a_s(\mu^2)}
      \right]^{\omega(n)}\, ,
\label{eq:Ediag}
\end{eqnarray}
\end{widetext}
where
$a_s(\mu^2)= \alpha_s^{\text{glob};(2)}(\mu^2,\Lambda_3^2)/(4\pi)$
and  $c_1=b_1/b_0$, with $b_i$ being the expansion coefficients of the
QCD $\beta$-function.
The evolution exponent of the coupling is defined by
$
 \omega(n)
=
 [\gamma_1(n)b_0-\gamma_0(n)b_1]/[2b_0b_1]
$.
The second term in the brackets in Eq.\ (\ref{eq:DA_EVO-glob})
represents the non-diagonal part of the evolution equation to the
order $O(a_s^2)$ induced by renormalization and encodes the mixing
of the higher Gegenbauer harmonics for indices $k > n$ related to
the conformal-symmetry breaking at NLO \cite{Mueller:1994cn}.
Notice that all components on the right-hand side of
Eqs.\ (\ref{eq:DA_EVO-glob}) and (\ref{eq:Ediag}) depend on $N_f$,
which changes to $N_f+1$, when the next quark-mass threshold is
crossed.
The explicit form of the mixing coefficients is given by
\cite{Bakulev:2002uc}
\begin{widetext}
\begin{eqnarray}
 d_{nk}(Q^2,\mu^2)
  & = &
  \frac{M_{nk}}{\gamma_0(k)-\gamma_0(n)-2b_0}
  \left\{1
    - \left[\frac{a_s(Q^2)}
                 {a_s(\mu^2)}
      \right]^{[\gamma_0(k)-\gamma_0(n)]/(2b_0)-1}
  \right\}\, ,
\end{eqnarray}
where the values of the first few elements
of the matrix $M_{nk}$ ($k=2, 4\geq n=0, 2$) read
\begin{eqnarray}
\label{eq:App-Mnk}
  M_{0 2} = -11.2 + 1.73  N_f, ~~
  M_{0 4} = -1.41 + 0.565 N_f, ~~
  M_{2 4} = -22.0 + 1.65  N_f\, .
\end{eqnarray}
\end{widetext}
Analytic expressions for $M_{nk}$ have been obtained
in \cite{Mueller:1993hg}.
The values in Eq.\ (\ref{eq:App-Mnk}) reproduce the exact
results with a deviation less than about $1\%$.

To make our further exposition more compact, we make use of the
parameter vectors
$\mathbf{A}_{}(\mu^2)$ and $\bm{\Psi}(x)$
defined at the reference momentum scale $\mu^2$ as follows
\begin{subequations}
\begin{eqnarray}
  \mathbf{A}
&=&
  (1,\,a_2,\,a_4\,,\cdots,\,a_{2(N-1)})\, ,
\\
  \bm{\Psi}
&=&
  (\psi_0,\psi_2,\cdots,\psi_{2(N-1)})\, ,
\\
  \varphi_{\pi}(x,\mu^2)
&=&
  \sum\limits_{n=0}^{N-1} a_{2n}(\mu^2)\psi_{2n}(x)
\nonumber \\
&=&
  \mathbf{A}_{}(\mu^2)\bm{\Psi}(x)\,,
\end{eqnarray}
\end{subequations}
where their dimension and the dimension of the matrix $E$ depends
on the parameter $N$.
Then, the evolution of the pion DA can be carried out in terms of the
Gegenbauer coefficients $a_i$ with $i=2, 4,\ldots ,2(N-1)$.
For a fixed number of flavors, one gets
\begin{subequations}
\begin{eqnarray}
  \bm{\Psi}(x;\mu^2)
& = &
  E(N_f,\mu^2,\mu^2_0)\bm{\Psi}(x), \\
\mathbf{A}(\mu^2)
& = &
  E^\text{T}(N_f,\mu^2,\mu^2_0)\mathbf{A}(\mu_0^2)\, ,
\end{eqnarray}
\end{subequations}
where $E^\text{T}$ is the transposed matrix of $E$, while
$\mathbf{A}(\mu_0^2)$ is the vector of the Gegenbauer coefficients
defined at some initial scale $\mu_0^2$.

In the global QCD scheme, the evolution of the pion DA defined at the
initial scale $\mu_0^2$, is implemented by means of the threshold interval
factors $E_i$ in the following step-by-step procedure,
\begin{widetext}
\begin{eqnarray}
  E_\text{glob}(\mu^2,\mu^2_0)
&=&
   E_3(\mu^2)\theta(\mu^2<M_4^2)
  +
   E_4(\mu^2)\theta(M_4^2\leqslant\mu^2<M_5^2)E_3
  + \nonumber \\
&&
   E_5(\mu^2)\theta(M_5^2\leqslant\mu^2<M_6^2)E_4E_3
  +
   E_6(\mu^2)\theta(M_6^2\leqslant\mu^2)E_5E_4E_3 \, ,
   \label{eq:global-pionDA-evo}
\end{eqnarray}
where the matrices $E_i$ and $E_i(\mu^2)$ are given by
\begin{eqnarray}
  E_i(\mu^2)
\equiv
  E(i,\mu^2,M^2_i) \, ,~~
  E_i
\equiv
  E(i,M^2_{i+1},M^2_i)
\end{eqnarray}
\end{widetext}
and the thresholds are defined \cite{Bakulev:2012sm} by the heavy-quark
masses
$m_c\sim M_4=1.65$~GeV,
$m_b \sim M_5=4.75$~GeV,
and $m_t\sim M_6 = 172.5$~,
while
$M^2_3\equiv \mu^2_0$ sets the initial scale
taken to be either $\mu_{0}=\mu_{1}=1$~GeV or $\mu_{0}=\mu_2=2$~GeV,
see Table \ref{tab:pion-DAs-a6}.
Note that the global evolution matrix,
Eq.\ (\ref{eq:global-pionDA-evo}),
is presented for $\mu_0<M_4$ and $\mu>\mu_0$.
No \textit{matching} at the mass thresholds is needed in the case of
equal initial and final momentum scales, i.e., $E(N_f;Q^2,Q^2)=1$
because of the independence of the evolution matrix on the number
of flavors.
For example, at the threshold $M_4$, we have $E_4(M_4^2)=1$
ensuring the continuity of the global evolution matrix $E_\text{glob}$.
It is worth noting that our NLO evolution scheme in terms
of Eq.\ (\ref{eq:global-pionDA-evo}),
has the following improvements relative to that used in
\cite{Bakulev:2004cu}
(see Appendix D there):
(a)	It is applicable to DAs with any number of Gegenbauer harmonics.
(b) The number of heavy-quark thresholds is extended to four flavors.
(c) When the interval of evolution contains two or more thresholds,
    our method can still incorporate contributions from the
    non-diagonal part of the evolution matrix
    removing the restriction to use only the first two Gegenbauer
    coefficients
    $a_2$ and $a_4$ as in \cite{Bakulev:2004cu}.

We reiterate that the matching of the coupling constants at the
quark-mass thresholds requires the readjustment of the value of the
QCD scale parameter
$\Lambda$ to $\Lambda_{(N_f)}$.
A detailed description of the matching procedure of the
running coupling in the global scheme can be found in
\cite{Bakulev:2012sm} and references cited therein.
For definiteness, we quote here the two-loop $\Lambda_{(N_f)}^{(2)}$ values
used in our code:
$\Lambda_{(3)}^{(2)} = 369$~MeV,
$\Lambda_{(4)}^{(2)} = 305$~MeV,
$\Lambda_{(5)}^{(2)} = 211$~MeV,
$\Lambda_{(6)}^{(2)} = 88$~MeV.
These values are defined by fixing the strong coupling
\begin{equation}
  \alpha_S(M_Z^2)=0.118
\end{equation}
at the scale of the Z boson mass $M_Z=91$~GeV.

We emphasize that for self-consistency reasons, the
global two-loop coupling
$\alpha_s^{\text{glob};(2)}(\mu^2,\Lambda_3)/(4\pi)$
should be used in \textit{all} functions entering
Eq.\ (\ref{eq:DA_EVO-glob}) that depend on the coupling with a variable
flavor number $N_f$.
Finally, the global evolution of the Gegenbauer coefficients is given
by
\begin{eqnarray}
  \mathbf{A}_\text{glob}(\mu^2)
=
  E^\text{T}_\text{glob}(\mu^2,\mu^2_0)\mathbf{A}(\mu_0^2)\, ,
\end{eqnarray}
whereas the global evolution of the pion DA assumes the form
\begin{widetext}
\begin{eqnarray}
  \varphi_{\pi}^\text{glob}(x,\mu^2)
&=&
  \mathbf{A}_{\text{glob}}(\mu^2)\bm{\Psi}(x)= \bm{A}(\mu_0^2)\bm{\Psi}(x;\mu^2)
=
  \sum\limits_{n=0}^{N-1} a_{2n}^{\text{glob}}(\mu^2)\psi_{2n}(x)
\, .
\label{eq:DA_EVO-glob-final}
\end{eqnarray}
\end{widetext}

\section{Data collection versus theoretical predictions}
\label{sec:data-theory}
In this appendix, we collect in Table \ref{tab:ff-values-table}
all existing sets of experimental data on the pion-photon transition
form factor together with our theoretical predictions.
The results obtained with the BMS pion DA \cite{Bakulev:2001pa} differ
from those we reported before in \cite{Bakulev:2011rp}.
The differences originate from the fact that we are using here an
updated theoretical framework; see the text for explanations and
\cite{Mikhailov:2016klg} for details.
In addition, we use the evolution scheme described in the previous
Appendix.
The numbers given in parentheses are new predictions calculated with
the platykurtic pion DA determined in \cite{Stefanis:2014nla}.
The displayed theoretical uncertainties for the BMS set of pion DAs
greatly overlap with those related to the platykurtic ones.
Therefore, no error bars for the latter have been displayed.

\begin{table*}[]
\begin{center}
\begin{ruledtabular}
\caption{Compilation of all existing data on the scaled TFF
$
 \tilde{Q}^2F^{\gamma^{*}\gamma\pi^0}(\tilde{Q}^2)
\equiv
 \mathcal{F}_{\gamma\pi}(\tilde{Q}^2)
$
from different measurements:
(i) CELLO \protect\cite{Behrend:1990sr},
(ii) CLEO \protect\cite{Gronberg:1997fj},
(iii) \textit{BABAR} \protect\cite{Aubert:2009mc},
and (iv) Belle \cite{Uehara:2012ag}.
The TFF is measured at $\tilde{Q}^2$ where the differential cross
sections assume their mean values computed by numerical integration.
The last column shows the theoretical predictions calculated in this
work at the same momentum value $\tilde{Q}^2$ for each bin
using as nonperturbative input the bimodal
BMS pion DA \protect\cite{Bakulev:2001pa} and taking into account the
chief theoretical uncertainties as explained in
Sec.\ \ref{sec:analysis-theory}.
The numbers in parentheses show the results obtained with the
platykurtic (pk) pion DA \protect\cite{Stefanis:2014nla} bearing
uncertainties inside the previous ones.
Both types of DAs have suppressed endpoint regions $x=0,1$
(see the text).
NLO evolution, starting at $\mu_1=1$~GeV
(Table \ref{tab:pion-DAs-a6}), and including heavy-quark
thresholds is employed, see App.\ \ref{sec:global-NLO-evolution}.
\label{tab:ff-values-table}}
\smallskip
\smallskip
\begin{tabular}{ccccccc}
 $Q^2$ {\small bin range}
        & $\tilde{Q}^2$
           & $\mathcal{F}_\text{CELLO}^{\gamma^*\gamma\pi^0}(\tilde{Q}^2)$
               & $\mathcal{F}_\text{CLEO}^{\gamma^*\gamma\pi^0}(\tilde{Q}^2)$
                      & $\mathcal{F}_\text{BABAR}^{\gamma^*\gamma\pi^0}(\tilde{Q}^2)$
                            & $\mathcal{F}_\text{Belle}^{\gamma^*\gamma\pi^0}(\tilde{Q}^2)$
                                    & $\mathcal{F}_{\rm BMS(pk)}^{\gamma^*\gamma\pi^0}(\tilde{Q}^2)$
\\ \phantom{.}
[GeV$^2$] & [GeV$^2$] & [0.01 $\times$ GeV] & [0.01 $\times$ GeV] & [0.01 $\times$ GeV] & [0.01 $\times$ GeV] & [0.01 $\times$ GeV]
\\
\hline               
 0.5 --  0.8 &  0.68 &  8.37$_{-0.73}^{+0.67}$ & --                       & --                    & --                             &  5.39$_{-3.37}^{+3.46}$(5.98)  \\
 0.8 --  1.1 &  0.94 &  9.58$_{-0.84}^{+0.78}$ & --                       & --                    & --                             &  7.70$_{-2.80}^{+2.90}$(7.95)  \\
 1.1 --  1.5 &  1.26 &  9.54$_{-1.12}^{+1.00}$ & --                       & --                    & --                             &  9.94$_{-2.50}^{+2.62}$(9.65)  \\
 1.5 --  1.8 &  1.64 & --                      & 12.1$\pm\,0.8\,\pm\,$0.3 & --                    & --                             & 11.78$_{-2.46}^{+2.6} $(11.05) \\
 1.5 --  2.1 &  1.70 & 12.08$_{-1.62}^{+1.43}$ & --                       & --                    & --                             & 12.00$_{-2.45}^{+2.59}$(11.23) \\
 1.8 --  2.0 &  1.90 & --                      & 11.7$\pm\,0.7\,\pm\,$0.3 & --                    & --                             & 12.66$_{-2.39}^{+2.55}$(11.75) \\
 2.0 --  2.2 &  2.10 & --                      & 13.8$\pm\,0.8\,\pm\,$0.3 & --                    & --                             & 13.18$_{-2.32}^{+2.49}$(12.19) \\
 2.1 --  2.7 &  2.17 & 16.43$_{-3.60}^{+2.94}$ & --                       & --                    & --                             & 13.33$_{-2.29}^{+2.46}$(12.33) \\
 2.2 --  2.4 &  2.30 & --                      & 12.7$\pm\,0.9\,\pm\,$0.3 & --                    & --                             & 13.59$_{-2.24}^{+2.42}$(12.56) \\
 2.4 --  2.6 &  2.50 & --                      & 13.5$\pm\,1.0\,\pm\,$0.3 & --                    & --                             & 13.93$_{-2.16}^{+2.34}$(12.88) \\
 2.6 --  2.8 &  2.70 & --                      & 15.1$\pm\,1.1\,\pm\,$0.4 & --                    & --                             & 14.20$_{-2.07}^{+2.25}$(13.16) \\
 2.8 --  3.1 &  2.94 & --                      & 13.7$\pm\,1.2\,\pm\,$0.3 & --                    & --                             & 14.46$_{-2.01}^{+2.18}$(13.43) \\
 3.1 --  3.5 &  3.29 & --                      & 14.5$\pm\,1.2\,\pm\,$0.4 & --                    & --                             & 14.75$_{-1.93}^{+2.09}$(13.76) \\
 3.5 --  4.0 &  3.74 & --                      & 13.2$\pm\,1.4\,\pm\,$0.3 & --                    & --                             & 15.01$_{-1.84}^{+1.99}$(14.10) \\
 4.0 --  4.5 &  4.24 & --                      & 13.4$\pm\,1.5\,\pm\,$0.3 & 15.04$\pm\,0.39$      & --                             & 15.21$_{-1.76}^{+1.88}$(14.39) \\
 4.0 --  5.0 &  4.46 & --                      & --                       & --                    & 12.9$\pm\,2.0\,\pm\,$0.6       & 15.28$_{-1.72}^{+1.84}$(14.50) \\
 4.5 --  5.0 &  4.74 & --                      & 15.4$\pm\,1.7\,\pm\,$0.4 & 14.91$\pm\,0.41$      & --                             & 15.36$_{-1.68}^{+1.79}$(14.62) \\
 5.0 --  5.5 &  5.24 & --                      & 14.5$\pm\,1.8\,\pm\,$0.4 & 15.74$\pm\,0.39$      & --                             & 15.48$_{-1.61}^{+1.71}$(14.81) \\
 5.0 --  6.0 &  5.47 & --                      & --                       & --                    & 14.0$\pm\,1.6\,\pm\,$0.7       & 15.52$_{-1.58}^{+1.68}$(14.89) \\
 5.5 --  6.0 &  5.74 & --                      & 15.5$\pm\,2.2\,\pm\,$0.4 & 15.60$\pm\,0.45$      & --                             & 15.57$_{-1.55}^{+1.64}$(14.97) \\
 6.0 --  7.0 &  6.47 & --                      & 14.8$\pm\,2.0\,\pm\,$0.4 & 16.35$\pm\,0.36$      & 16.1$\pm\,0.7\,\pm\,$0.8       & 15.68$_{-1.48}^{+1.56}$(15.15) \\
 7.0 --  8.0 &  7.47 & --                      & --                       & 16.06$\pm\,0.47$      & 15.8$\pm\,0.6\,\pm\,$0.7       & 15.80$_{-1.40}^{+1.47}$(15.35) \\
 7.0 --  9.0 &  7.90 & --                      & 16.7$\pm\,2.5\,\pm\,$0.4 & --                    & --                             & 15.84$_{-1.37}^{+1.44}$(15.42) \\
 8.0 --  9.0 &  8.48 & --                      & --                       & 16.73$\pm\,0.60$      & 17.5$\pm\,0.5\,\pm\,$0.7       & 15.89$_{-1.34}^{+1.39}$(15.51) \\
 9.0 -- 10.0 &  9.48 & --                      & --                       & 18.53$\pm\,0.55$      & 16.9$\pm\,0.5\,\pm\,$0.7       & 15.97$_{-1.28}^{+1.33}$(15.63) \\
10.0 -- 11.0 & 10.48 & --                      & --                       & 18.66$\pm\,0.76$      & 16.5$\pm\,0.6\,\pm\,$0.7       & 16.04$_{-1.24}^{+1.28}$(15.73) \\
11.0 -- 12.0 & 11.48 & --                      & --                       & --                    & 17.3$\pm\,0.8\,\pm\,$0.7       & 16.10$_{-1.20}^{+1.24}$(15.82) \\
11.0 -- 12.0 & 11.49 & --                      & --                       & 19.16$\pm\,0.78$      & --                             & 16.10$_{-1.20}^{+1.24}$(15.82) \\
12.0 -- 13.5 & 12.71 & --                      & --                       & 17.50$\pm\,1.10$      & --                             & 16.16$_{-1.16}^{+1.19}$(15.91) \\
12.0 -- 14.0 & 12.94 & --                      & --                       & --                    & 16.8$\pm\,0.7\,\pm\,$1.0       & 16.18$_{-1.15}^{+1.18}$(15.93) \\
13.5 -- 15.0 & 14.22 & --                      & --                       & 19.80$\pm\,1.20$      & --                             & 16.23$_{-1.12}^{+1.14}$(16.00) \\
14.0 -- 16.0 & 14.95 & --                      & --                       & --                    & 17.9$\pm\,1.2\,\pm\,$1.3       & 16.26$_{-1.10}^{+1.12}$(16.04) \\
15.0 -- 17.0 & 15.95 & --                      & --                       & 20.80$\pm\,1.20$      & --                             & 16.30$_{-1.08}^{+1.10}$(16.09) \\
16.0 -- 18.0 & 16.96 & --                      & --                       & --                    & 18.3$\pm\,1.7\,\pm\,$1.2       & 16.33$_{-1.06}^{+1.08}$(16.13) \\
17.0 -- 20.0 & 18.40 & --                      & --                       & 22.00$\pm\,1.30$      & --                             & 16.38$_{-1.03}^{+1.05}$(16.18) \\
18.0 -- 20.0 & 18.96 & --                      & --                       & --                    & 19.8$\pm\,1.9\,\pm\,$1.3       & 16.39$_{-1.02}^{+1.04}$(16.20) \\
20.0 -- 25.0 & 22.28 & --                      & --                       & 24.50$\pm\,1.80$      & --                             & 16.47$_{-0.97}^{+0.98}$(16.30) \\
20.0 -- 25.0 & 22.29 & --                      & --                       & --                    & 19.5$\pm\,1.7\,\pm\,$1.3       & 16.47$_{-0.97}^{+0.98}$(16.30) \\
25.0 -- 30.0 & 27.31 & --                      & --                       & 18.10$_{-4.0}^{+3.3}$ & --                             & 16.56$_{-0.92}^{+0.93}$(16.40) \\
25.0 -- 30.0 & 27.33 & --                      & --                       & --                    & 23.6$_{-2.9}^{+2.6}$$\pm\,$1.6 & 16.56$_{-0.92}^{+0.93}$(16.40) \\
30.0 -- 40.0 & 34.36 & --                      & --                       & 28.50$_{-4.5}^{+3.9}$ & --                             & 16.64$_{-0.87}^{+0.87}$(16.49) \\
30.0 -- 40.0 & 34.46 & --                      & --                       & --                    & 18.8$_{-4.3}^{+3.5}$$\pm\,$1.3 & 16.64$_{-0.87}^{+0.87}$(16.50) \\
\end{tabular}
\end{ruledtabular}
\end{center}
\end{table*}

It is instructive to make some important remarks concerning
the CELLO data reported in \cite{Behrend:1990sr}.
These data were presented for the quantity
$\frac{F^{2}M^3}{64\pi}~\text{eV} \equiv a$,
evaluated at the reference momentum scale
$\langle Q^2 \rangle \equiv \tilde{Q}^2$.
They have been converted here to the quantity $Q^2F(Q^2)$ using
the relation
$
 Q^2|F^{\gamma^*\gamma\pi^0}(\tilde{Q}^2)|
=
 \frac{1}{4\pi\alpha}
 \sqrt{\frac{64\pi a}{M^{3}}} |Q^2|~\text{GeV}
$,
where $M\simeq 135$~MeV and $\alpha=1/137$.

It is worth noting that the CELLO data are usually shown for the
quantity
$Q^2F^{\gamma*\gamma\pi^0}(Q^2)$
not at the scale $\tilde{Q}^2$ but rather at the symmetric point of
each $Q^2$ interval, i.e., at the scale
$\bar{Q^2}=(Q_\text{max}^2 + Q_\text{min}^2)/2$.
The resulting deviations of the scaled TFF
$\tilde{Q}^2F^{\gamma^{*}\gamma\pi^0}(\bar{Q}^2)$
from $\tilde{Q}^2F^{\gamma^{*}\gamma\pi^0}(\tilde{Q}^2)$
are very small at lower $Q^2$ but they increase
with $Q^2$, becoming strongest at the highest scale probed, viz.,
$\bar{Q^2}=2.40$~GeV$^2$ for which one has
$
 Q^2F^{\gamma^{*}\gamma\pi^0}(\bar{Q^2})[0.01 \times \text{GeV}]
=
 18.17_{-3.98}^{+3.25}
$
instead of
$
 Q^2F^{\gamma^{*}\gamma\pi^0}(\tilde{Q}^2)[0.01 \times \text{GeV}]
=
 16.43_{-3.60}^{+2.94}
$,
see Table \ref{tab:ff-values-table}.
\end{appendix}


\end{document}